\documentclass[11pt]{article}
\pdfoutput=1
\usepackage{jheppub}
\usepackage[squaren,Gray]{SIunits}
\usepackage{subfigure}

\title{Multi--tops at the LHC}
\author[a,1]{Aldo Deandrea,\note{also Institut Universitaire de France, 103 boulevard Saint-Michel, 75005 Paris, France}}
\author[a]{Nicolas Deutschmann}
\affiliation[a]{Universit\'e de Lyon, France; Universit\'e Lyon 1, CNRS/IN2P3, UMR5822 IPNL,\\ F-69622 Villeurbanne Cedex, France.}
\emailAdd{deandrea@ipnl.in2p3.fr}
\emailAdd{n.deutschmann@ipnl.in2p3.fr}

\abstract{The experiments at the LHC are searching for many different final states that can hint to the presence of new physics 
beyond the Standard Model. One of the most interesting and promising sectors for these searches is that of the
top quark, for both theoretical and phenomenological reasons linked to its large mass and to its possible special role in the 
electroweak symmetry breaking sector. We suggest that multi-top events, beyond the standard $t$-$\bar t$ and four top searches, 
can bring further insight in constraining and discovering physics beyond the Standard Model, taking advantage of experimental 
techniques similar to those used in present top-quark analyses. This is relevant both for the next data taking runs at the LHC and even 
more at higher luminosity and higher energy collider options, which are discussed for future LHC upgrades and future accelerators. 
In particular we consider six top and eight top final states, discussing the generic colour representations for beyond the Standard Model 
particles giving rise to those final state. 
We also discuss the limits which can be extracted by using the present analyses sensitive to four top final states, as well as
the potential bounds from new searches we propose to experimental collaborations as an alternative.}
\preprint{LYCEN 2014-06} 
\begin{document}
\maketitle

\section{Introduction: closing the window on top multiplicity}

The top quark holds a special place in the Standard Model (SM) and in its extensions. Indeed it is one of the central points of interest 
for the experimental collaborations at the LHC due to its mass and its peculiar properties compared to the other quarks on the 
phenomenological side, and due to its link to the electroweak sector and to physics beyond the SM on the theoretical side.

One very interesting question is whether present searches like those for four top final states can be extended to multi-top final states
with more tops, and in particular what can be learned in this exercise. It turns out that even if one can think naively that a very large 
number of top particles can be produced at the LHC, these multi-top final states in practice have many constraints both regarding the 
number of particles and the type of production processes. In the following we only perform a preliminary study and we consider 
simple generic models that yield large multiplicities of top quarks in their final states. A detailed study is not possible without considering 
all the experimental details, therefore this note is meant only as a suggestion for further analyses by the LHC experimental 
collaborations. At present only few pioneering experimental analyses have started to focus on related final states (see for
example \cite{CMS:2012nca} which considers an 8-jet final state), but one could extend these analyses to top quarks.
\begin{figure}[!ht]
\centering
\includegraphics[scale=0.8,trim = 15mm 150mm 0mm 20mm]{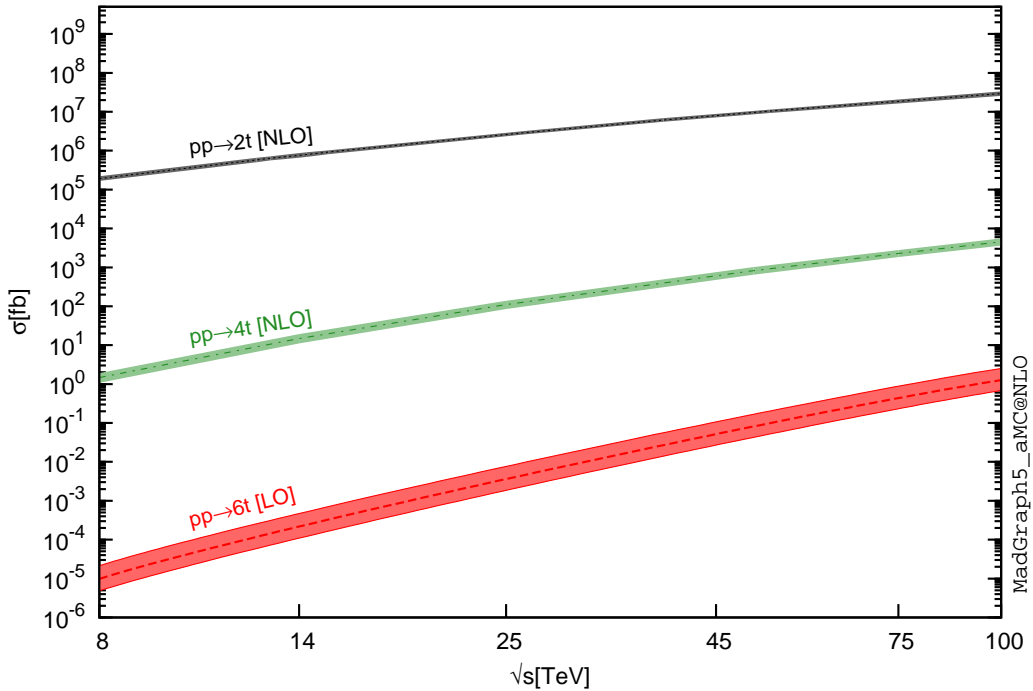}
\vspace{2em}
\caption{Multitop production cross-sections in the Standard Model as a function of the centre of mass energy for the colliding protons. 
The simulations were carried out using Madgraph5\_aMC@NLO \cite{Alwall:2014hca} and the error bands reflect the scale uncertainty.}
\label{plotsm}
\end{figure}
As the number of particles grows, the size of the available phase space shrinks and one can guess that there is a limit
on the number of top quarks which can be produced in a single observable event. 
In the Standard Model, this number however lies well below the naive estimate
$N_{\mathrm{max}}=\dfrac{\sqrt{s}}{m_t}\approx 80$ for
$\sqrt{s}=\unit{14}{\tera\electronvolt}$, and as Figure~\ref{plotsm}
shows, there is little hope to see more than 4 tops at the LHC. Even beyond the Standard Model the maximal top
multiplicity observable at the LHC stays much smaller than $N_{\mathrm{max}}$.

In this note we attempt to give an idea of this limit using a set of simple and generic effective models which yield six or more top 
quarks in the final state. 

\section{Toy models for multi-top physics}

If several different models can give rise to the final states we are interested in, we limit ourselves to a set of models based on the
assumption that new physics couples only to top quarks (a kind of ``top portal''). The topologies we consider, consist of the decay 
chains of pair-produced coloured particles, which will either be ``coloron''-like bosons (see for example \cite{Simmons:2013zoa}) 
or $t'$-like vector-like fermions (see for example \cite{Cacciapaglia:2010vn,Okada:2012gy,Buchkremer:2013bha}).

It is important to note that no matter how complicated the decay chain, there is only one parameter that strongly influences the event 
yield: the mass of the original pair-produced particle, as its production mechanism is fixed by QCD and we assume a unique decay 
channel. Hence limits and observation windows estimated using cuts and simple event 
counting with these topologies are fairly generic and can be generalised to many other models with the same signatures.

\section{Six tops}
As the four top final state is already extensively studied in the 
literature \cite{Battaglia:2010xq,Gregoire:2011ka,Cacciapaglia:2011kz,AguilarSaavedra:2011ck}, and searched in detail by the LHC
experimental collaborations \cite{ATLAS:2012hpa,CMS:2013xma}, 
we start considering the six top quark final state, by analysing
the production and decay of such a state together with simple analyses which can be used to bound this process.

\subsection{Model and production process}
\begin{figure}[!ht]
\centering
\includegraphics{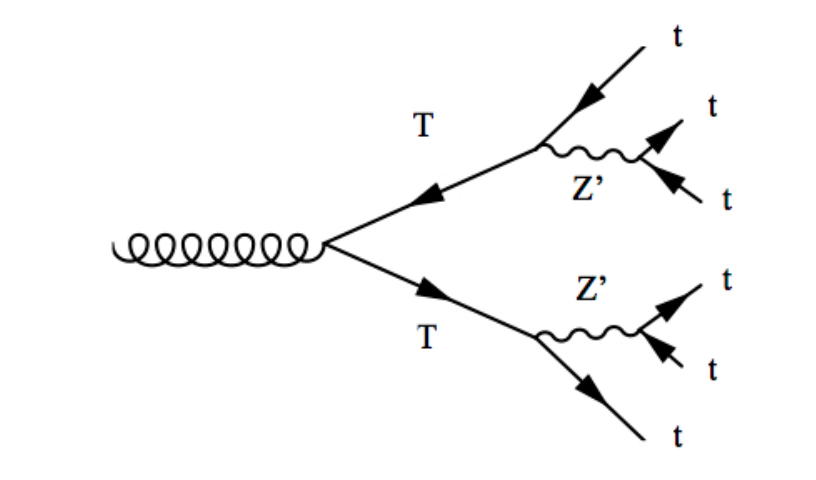}
\caption{Six top production mechanism}
\label{fig:tttbar}
\end{figure}

Reaching a six top final state requires two particles beyond the Standard Model: a top partner $T$  with mass $M_T$ and a bosonic
particle which we will call $Z'$, with mass $M_{Z'}$. The six top final state is produced in the process shown in Figure~\ref{fig:tttbar}.

The most natural gauge group representation is $(\mathbf{3},\mathbf{1},2/3)$ for $T$ and $(\mathbf{1},\mathbf{1},0)$ for $Z'$. 
More exotic colour embeddings are possible, as listed in Table \ref{tab:colorembed6t}, but having an additional coloured particle 
obviously leads to a richer --- and hence more constrained --- model. In the following
subsections, we will set limits on this model using searches for new
physics at the LHC, which could be sensitive to the final state of Figure~\ref{fig:tttbar}.
\begin{table}[!ht]
\centering
\begin{tabular}{|c|c|c|}
\hline
 &$R_{Z'}$&$R_T$ \\\hline
$R_1$ & $\mathbf{1}$&$\mathbf{3}$\\\hline
$R_2$ & $\mathbf{8}$&$\mathbf{3} $\\\hline
$R_3$ & $\mathbf{8}$&$\bar{\mathbf{6}} $\\
\hline
$R_4$ & $\mathbf{8}$&$\mathbf{15}$\\
\hline
\end{tabular}
\caption{All possible colour embeddings for $T$ and $Z'$ in the topology of Figure \protect{\ref{fig:tttbar}}}
  \label{tab:colorembed6t}
\end{table}

\subsection{Kinematic Distributions}
\begin{figure}[!ht]
\centering
\subfigure[$H_T$ with $M_T=\unit{720}{\giga\electronvolt}$ (blue crosses) and
$M_T=\unit{1000}{\giga\electronvolt}$ (purple discs) at $\sqrt{s}=\unit{8}{\tera\electronvolt}$]{\includegraphics[scale=0.3]{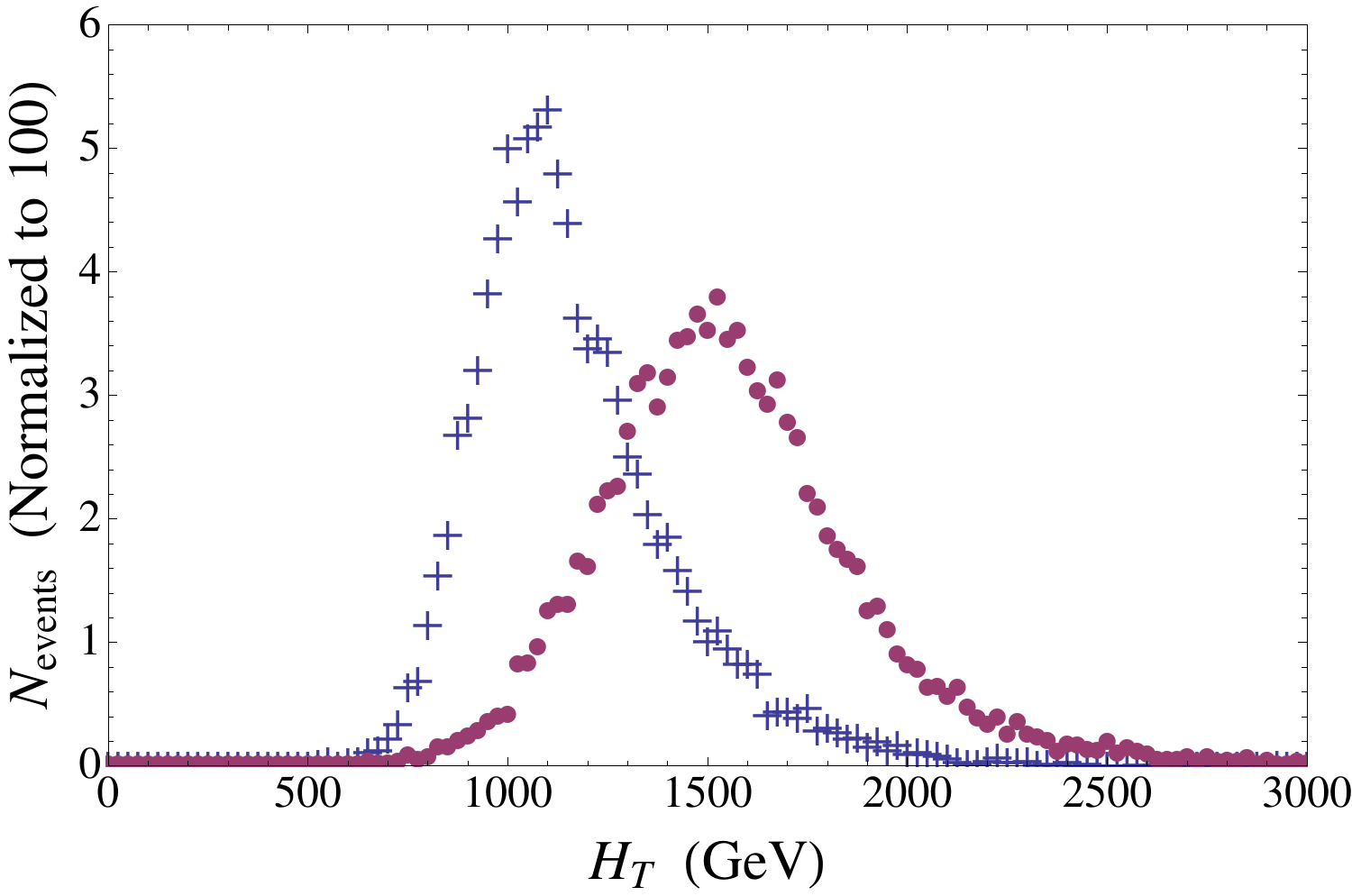} \label{a}}\hspace{1em}
\subfigure[$H_T$ with $M_T=\unit{720}{\giga\electronvolt}$ at
$\sqrt{s}=\unit{8}{\tera\electronvolt}$
 (blue crosses) and
$\sqrt{s}=\unit{14}{\tera\electronvolt}$ (purple discs)]{\includegraphics[scale=0.3]{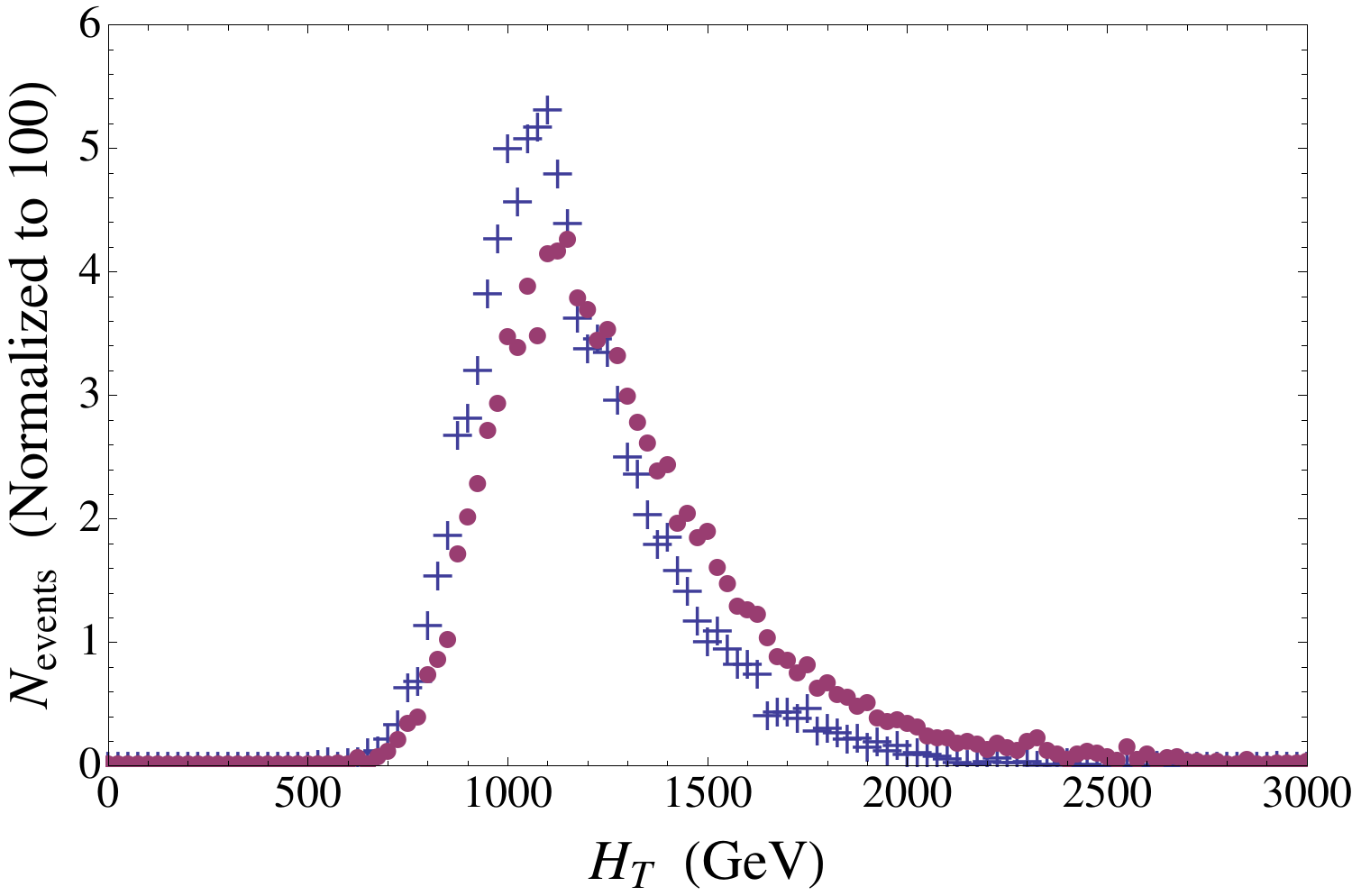} \label{b}}\hspace{1em}
\subfigure[${\not}E_T$ with $M_T=\unit{720}{\giga\electronvolt}$ (blue crosses) and
$M_T=\unit{1000}{\giga\electronvolt}$ (purple discs) at $\sqrt{s}=\unit{8}{\tera\electronvolt}$]{\includegraphics[scale=0.3]{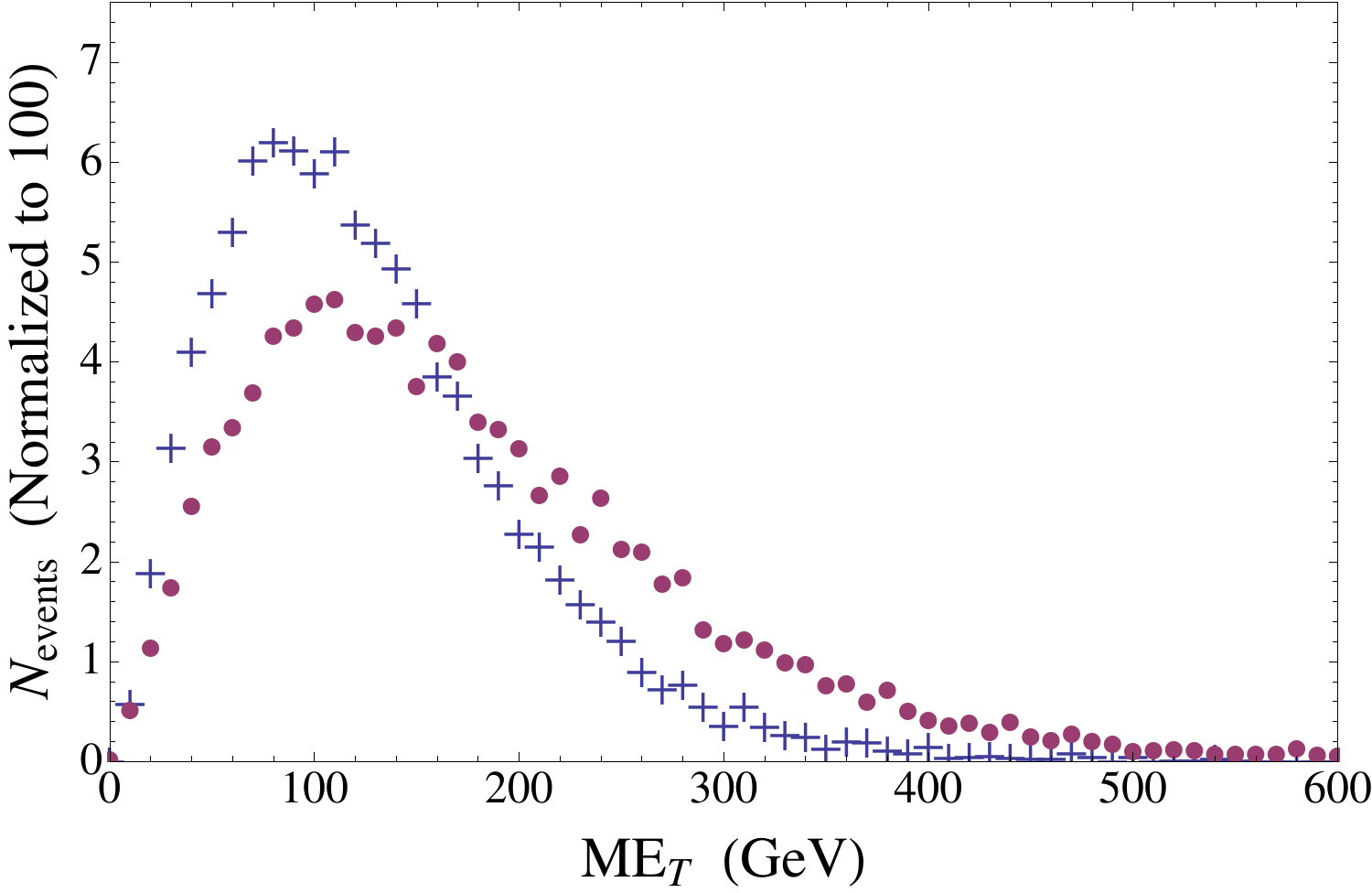} \label{c}}

\subfigure[${\not}E_T$ with $M_T=\unit{720}{\giga\electronvolt}$ at
$\sqrt{s}=\unit{8}{\tera\electronvolt}$ (blue crosses) and
$\sqrt{s}=\unit{14}{\tera\electronvolt}$ (purple discs)]{\includegraphics[scale=0.3]{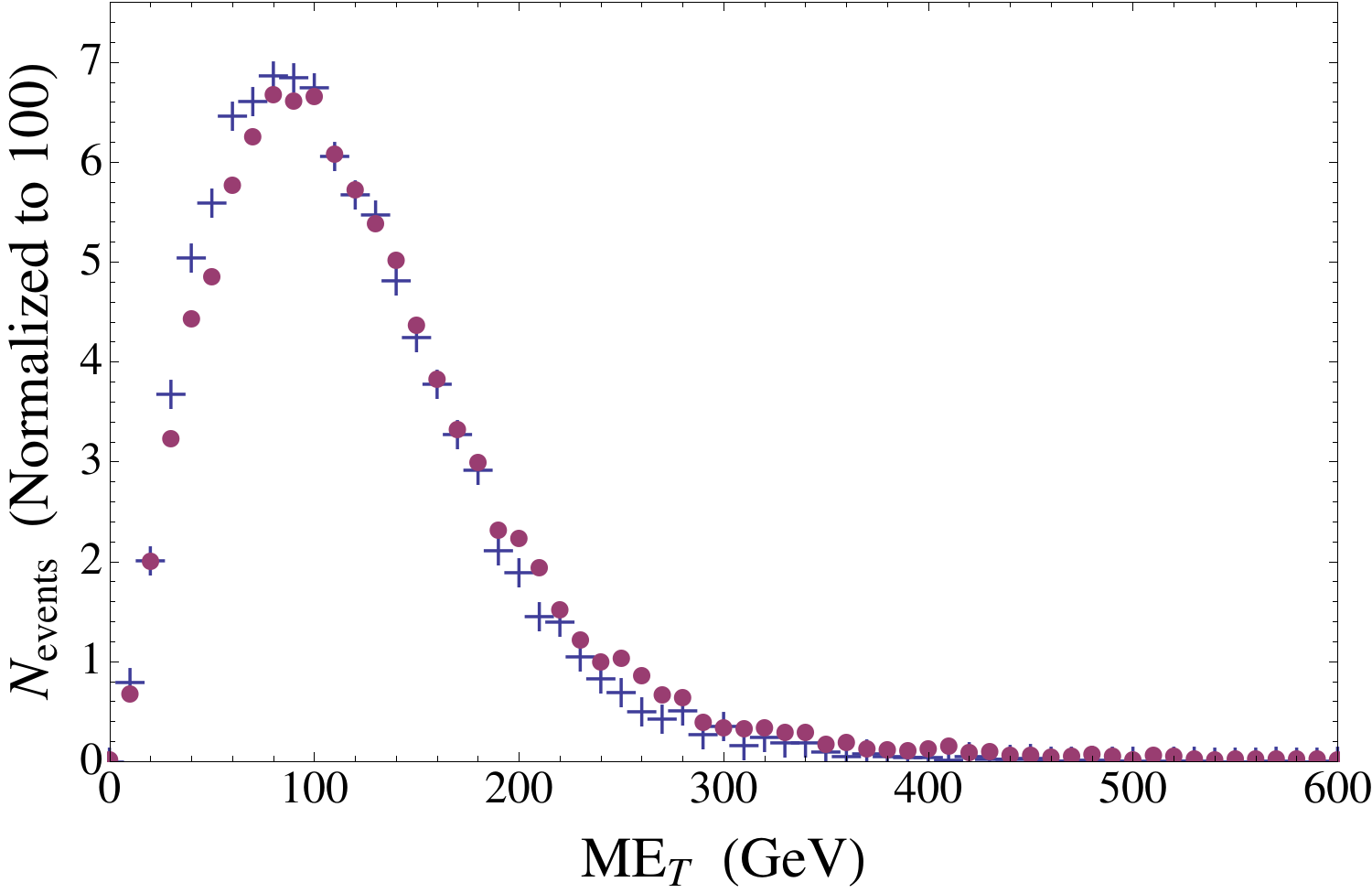} \label{d}}
\hspace{1em}
\subfigure[$p_T$ of the first lepton with $M_T=\unit{720}{\giga\electronvolt}$ (blue crosses) and
$M_T=\unit{1000}{\giga\electronvolt}$ (purple discs) at $\sqrt{s}=\unit{8}{\tera\electronvolt}$]{\includegraphics[scale=0.3]{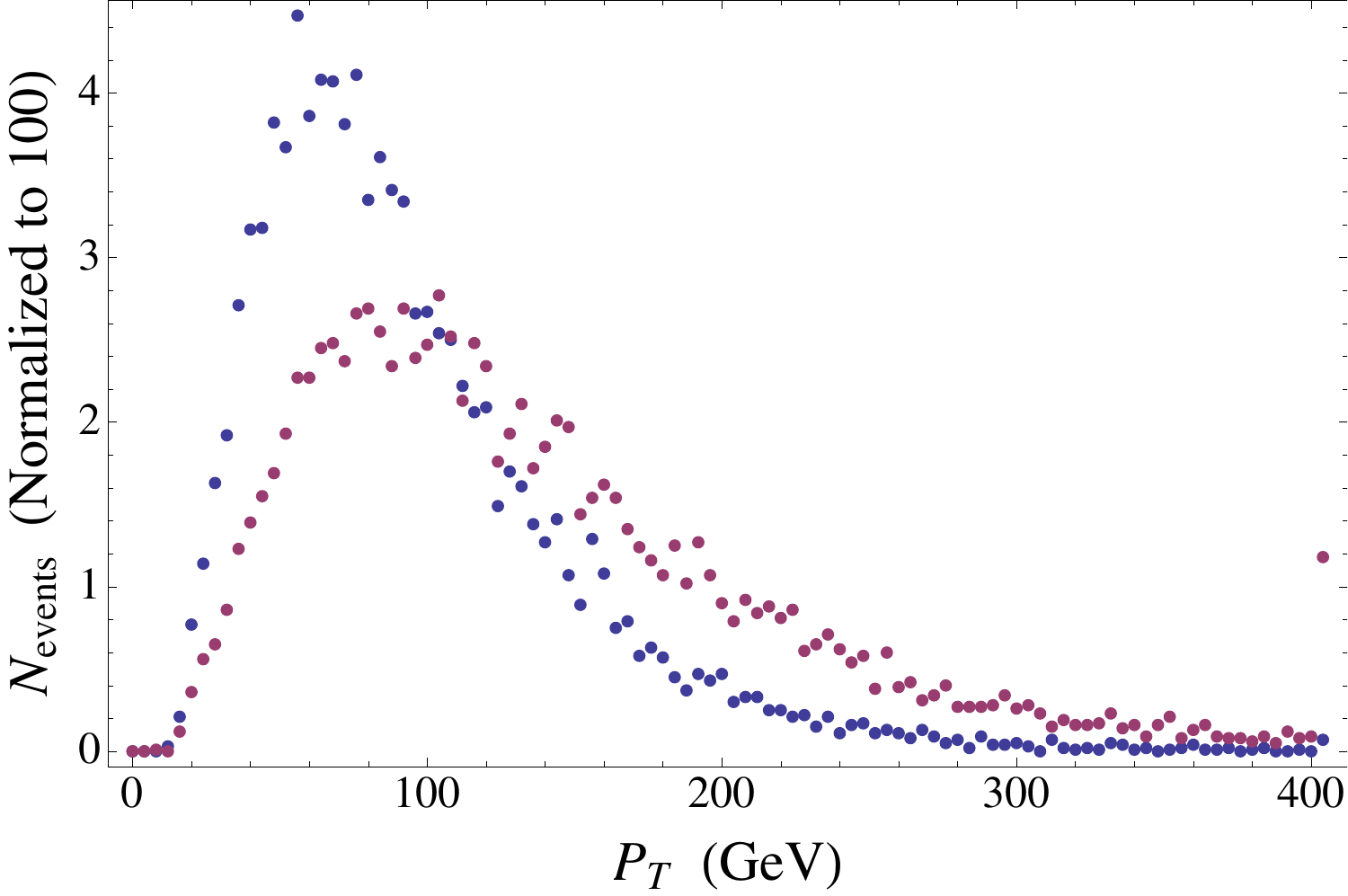} \label{e}}\hspace{1em}
\subfigure[$p_T$ of the first lepton with $M_T=\unit{720}{\giga\electronvolt}$ at
$\sqrt{s}=\unit{8}{\tera\electronvolt}$ (blue crosses) and
$\sqrt{s}=\unit{14}{\tera\electronvolt}$ (purple discs)]{\includegraphics[scale=0.3]{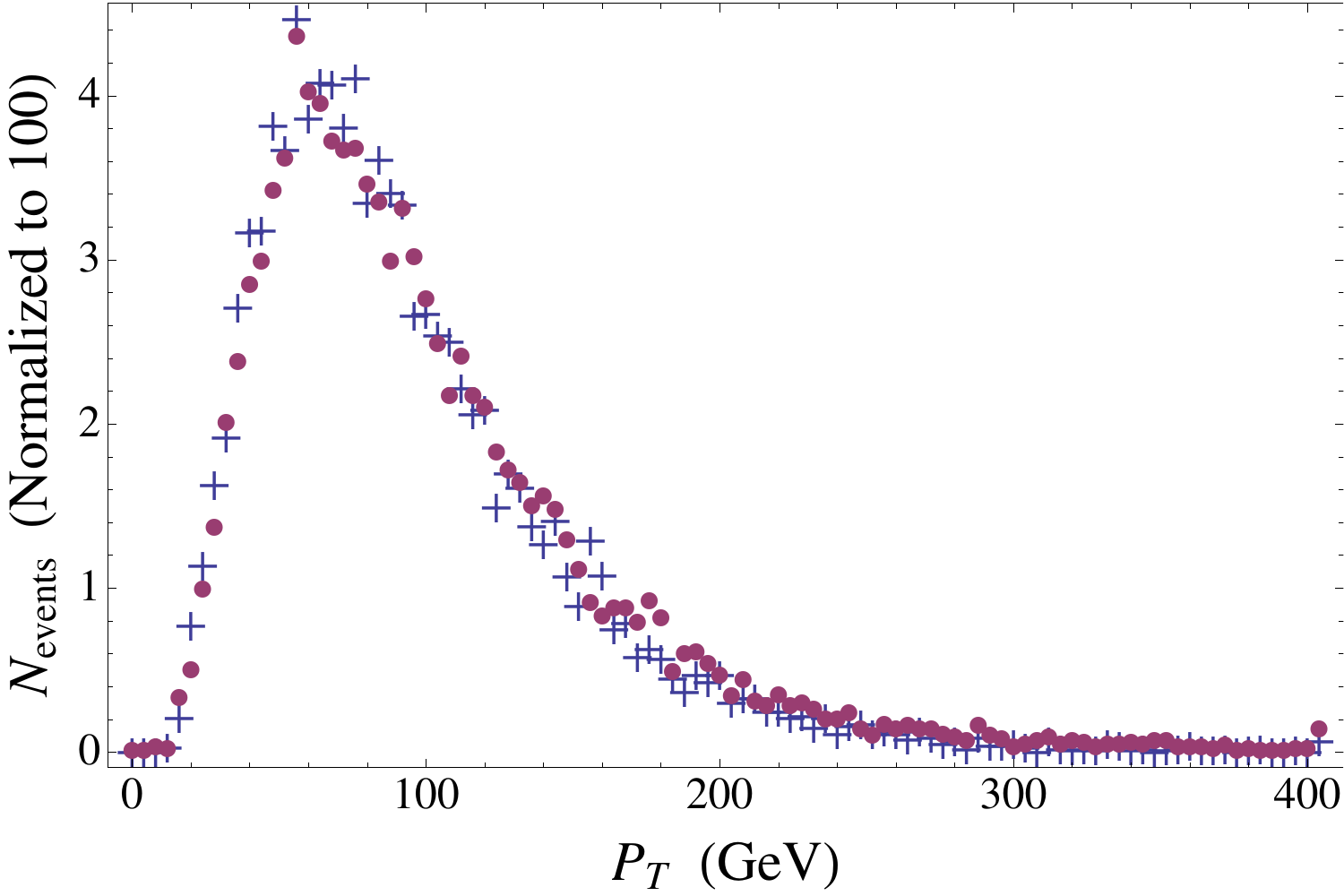} \label{f}}
\subfigure[$\Delta R $ between the first lepton and jets (multiple
entries per event) with $M_T=\unit{720}{\giga\electronvolt}$ (blue crosses) and
$M_T=\unit{1000}{\giga\electronvolt}$ (purple discs) at
$\sqrt{s}=\unit{14}{\tera\electronvolt}$]{\includegraphics[scale=0.3]{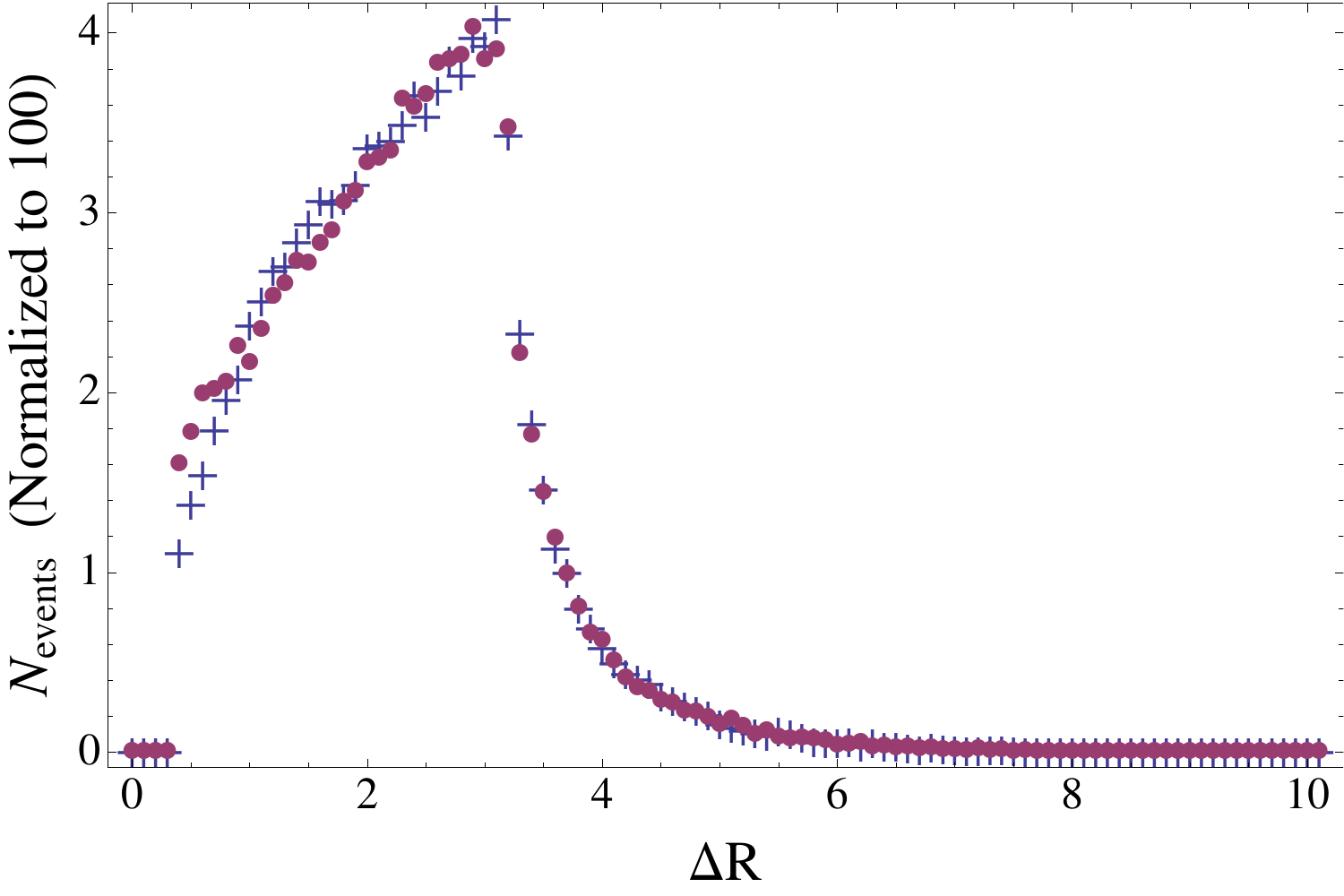} \label{g}}\hspace{1em}
\subfigure[$\Delta R $ between the first lepton and jets with $M_T=\unit{720}{\giga\electronvolt}$ (blue crosses) and
$M_T=\unit{1000}{\giga\electronvolt}$ (purple discs) at $\sqrt{s}=\unit{8}{\tera\electronvolt}$]{\includegraphics[scale=0.3]{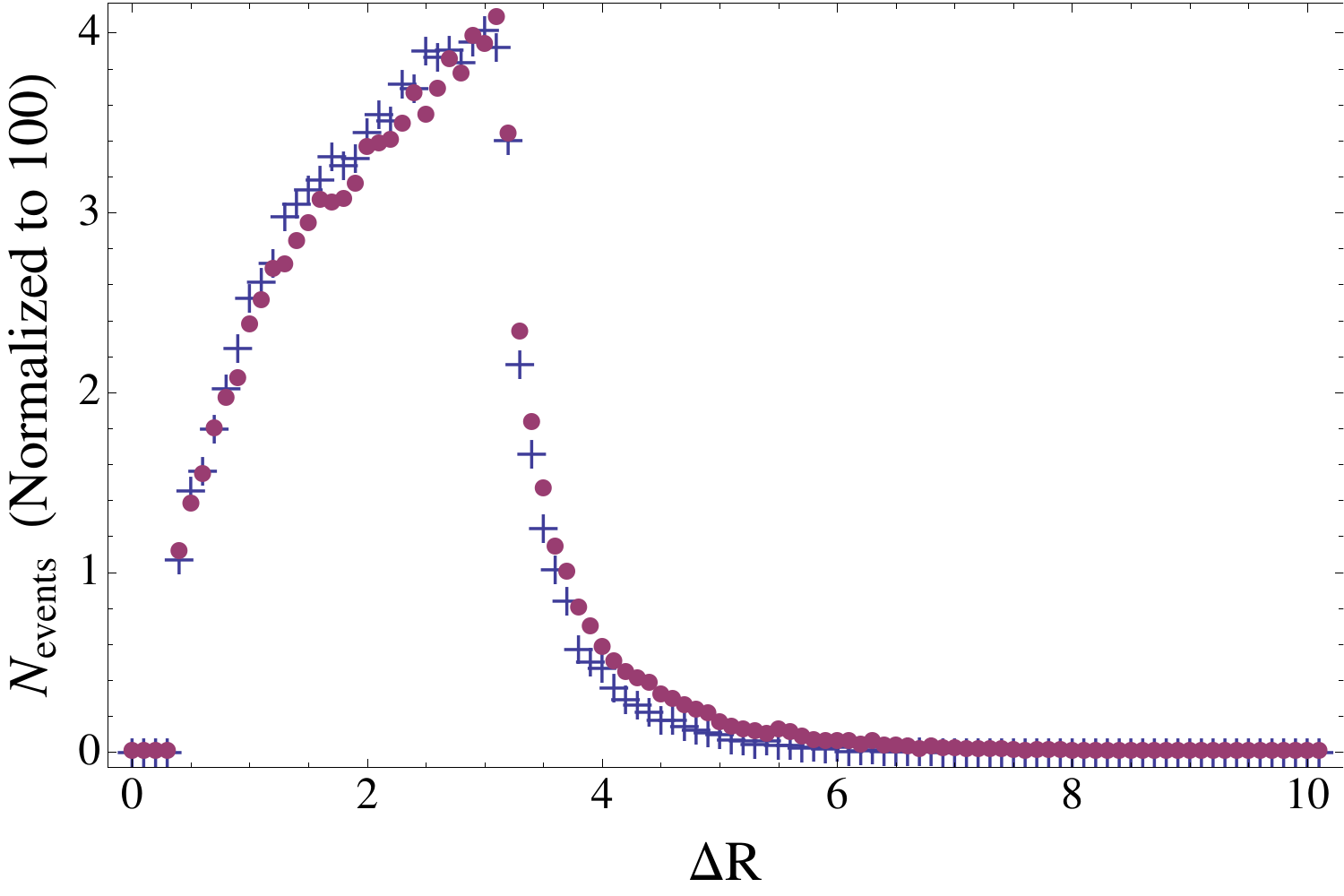} \label{h}}
\caption{Kinematic distributions of the six-top process.}
\end{figure}

As usual with models involving a high-mass new particle, the process
described above always involves a high total transverse energy $H_T$. The bulk of the distribution is shifted to higher energies
as $M_T$ grows (Figure~\ref{a}), and so with $\sqrt{s}$, in a milder fashion (Figure~\ref{b}). But even at the
kinematic limit and $\unit{8}{\tera\electronvolt}$, it is largely sufficient for a large majority of events to pass the most selective
QCD-background-reducing cuts used in CMS lepton analyses (see \cite{Chatrchyan:2012paa}). Since we
focus on events where two tops decay leptonically, there is also a
large amount of ${\not}E_T$ coming from the undetected neutrinos, which again is
typically sufficient to be distinguished from QCD events. The shape of
the ${\not}E_T$ distribution has very low dependence in the centre-of-mass energy of the collisions
(Figure~\ref{d}), which is also true for the transverse momentum of observed particles, such as
leptons and jets (Figure~\ref{f}). The shapes of these variable do however depend on $M_T$  (Figures \ref{c} and \ref{e}). 
Angular distance variables show the same lack of
$\sqrt{s}$-dependence (Figure~\ref{h}) and also do not depend on $M_T$ (Figure~\ref{g}). This shows that
due to the high multiplicity of the event, little correlation is kept between the leptons and the other decay products and one can 
consider the events as almost spherical.

\subsection{Existing limits}
In the following section we examine the present bounds on the six--top final state coming from different analyses, mainly those 
including two and three same sign leptons, which have a very reduced
background. We also discuss simple improvements in these search strategies.

\subsubsection{Two same-sign leptons with b-jets}
Among the large possible number of detector--level signatures of the six--top
final state, those involving several same-sign leptons seem the
most promising. These leptons will be accompanied by a large number
of b-jets and jets. The CMS search for same-sign dileptons production associated with b-jets (2SSL+b) presented in
\cite{Chatrchyan:2012paa} is sensitive to such final states, so we
reproduced their analysis on simulated events from our model to set
limits on our parameter space. We specialised to the signal region 7 of the search
(SR7), whose selection criteria are detailed in Table \ref{tab:kin},
due to its low background expectation value. 

\begin{table}[!ht]
\centering
\begin{tabular}{|l|c|}\hline
Variable              & Cut \\ \hline
\# leptons charges    & $++$ or $--$\\
\# jets               & $\geq 3$\\
\# b-tags             & $\geq 3$\\
$H_T$                 & $> \unit{200}{\giga\electronvolt}$\\
${\not}E_T$           & $> \unit{50}{\giga\electronvolt}$\\
lepton $p_T$          & $> \unit{20}{\giga\electronvolt}$ \\
$\sum p_T$ within $\Delta R \leq 0.3$ of lepton & $\leq 10\%$ of lepton's $p_T$\\
electronic $\eta$     &  $|\eta| < 1.442$ or $1.566 < |\eta| < 2.4$ \\
muonic $\eta$          & $|\eta| < 2.4$ \\
jets $p_T$            & $p_T > \unit{40}{\giga\electronvolt}$ \\
jets $\eta$           & $|\eta| < 2.4$ \\\hline
\end{tabular}
\caption{\label{tab:kin} Kinematic requirements for this analysis. }
\end{table}

Other analyses could be sensitive to these multi-lepton, multi-jet, ${\not}E_T$ final states, in particular ATLAS supersymmetry searches in the squark-gluino searches \cite{Aad:2014qaa} and \cite{Aad:2014pda}. However they both have drawbacks that make \cite{Chatrchyan:2012paa} seem most relevant: the first only focuses on opposite sign dileptons and hence will have a larger background and put weaker limits while in the latter, only signal regions SR3b and SR3L would be sensitive due to the chargino invariant mass cut of the others. However, SR3b does not use ${\not}E_T$ in its selection which allows for more QCD background and SR3L does not use b-tagging information. 

The events were generated by MadGraph 5 \cite{Alwall:2011uj} and further hadronised in Pythia 6 \cite{Sjostrand:2006za}. The jets
were reconstructed using FastJet \cite{Cacciari:2011ma} with the anti-$k_T$ algorithm ($R=0.5$ as used in the original search on 
real data) and we applied the cuts in Table \ref{tab:kin} to the reconstructed-level
variables with MadAnalysis 5 \cite{Conte:2012fm}. This provides an
event yield expectation value for our signal, for which we can compute
a confidence level for the signal to be compatible with the observed
number of events using the $CL_s$ method \cite{Junk:1999kv}. The expected background
and measured number of event are taken from the CMS search and both signal and
background are modelled using Poisson statistics. Of course, we apply
our selection at the reconstructed level without proper detector
simulation, which means that our confidence levels should be taken for
what they are worth: gross estimates for a preliminary exploration of
multitop final states. 

\begin{figure}[!ht]
\centering
\subfigure[]{\includegraphics[scale=0.55]{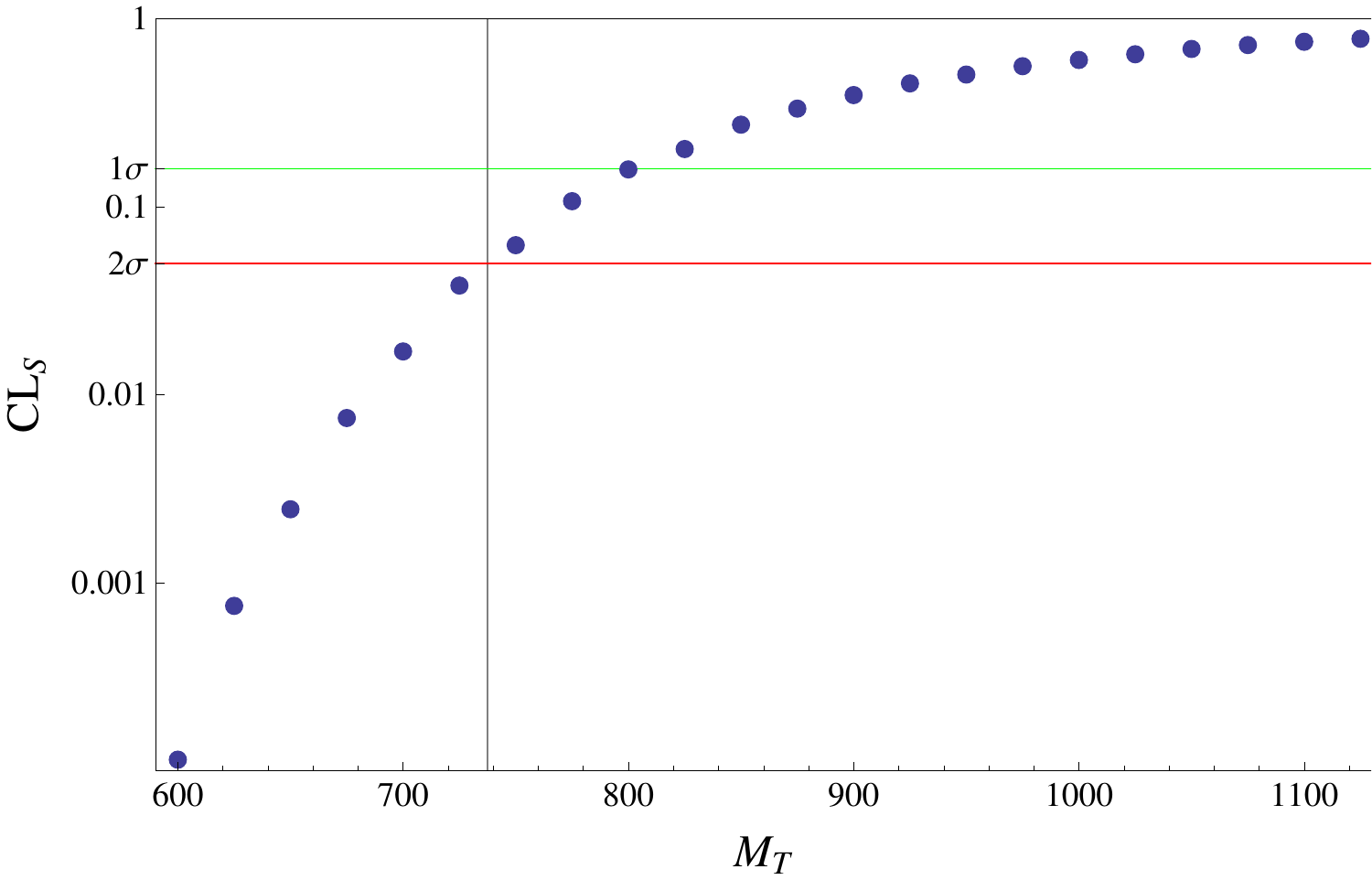} \label{limmtmzb}}\hspace{1em}
\subfigure[]{\includegraphics[scale=0.55]{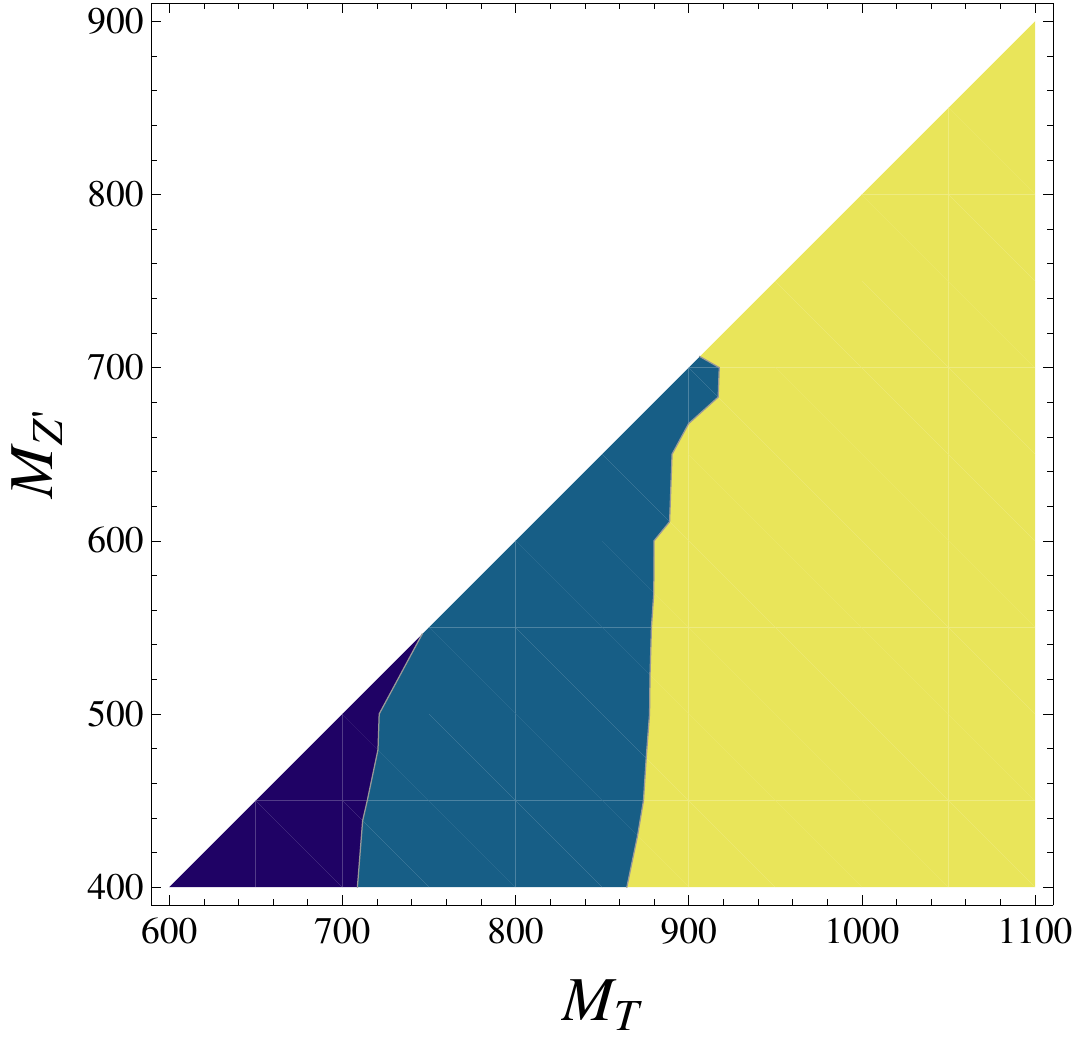} \label{limmtmza}}
\caption{\subref{limmtmzb}: $CL_s$ confidence level in a scan over $M_T$ with
$M_{Z'}=\unit{400}{\giga\electronvolt}$. \subref{limmtmza}: Limits in the plane $(M_T,M_{Z'})$ from CMS. Purple points
  are excluded at $2\sigma$, blue points are excluded at $1\sigma$ and
yellow points are not excluded.}
\label{limmtmz}
\end{figure}

We performed a parameter scan in the plane $M_T,\ M_{Z'}$ in the minimal colour embedding model and wehave been able to exclude 
the region below $M_T<\unit{710}{\giga\electronvolt}$, as shown in Figure~\ref{limmtmz}. As expected the limit has very mild
dependance on $M_{Z'}$ so we will keep it at $\unit{400}{\giga\electronvolt}$ and only scan over $M_T$ for further discussions.

We did not account for K-factors in our limits, which means that we underestimate the parton-level cross-sections and thus the limit on $M_T$. This is justified by comparing $t\overline{t}$ production in Madgraph5\_aMC@NLO at LO and NLO with $m_t=\unit{600-1000}{\giga\electronvolt}$ which yields a $<20\%$ increase in the overall cross-section. Such a factor would change unsignificantly the limit we set on $M_T$.

This limit could be extrapolated to the more unusual colour structures
of Table \ref{tab:colorembed6t}  by multiplication of the signal yield by
adapting the color factor in the approximation where we neglect colour correlation effects. 

\subsubsection{Possible improvements in the event selection}

\begin{figure}[!ht]
\centering
\includegraphics[width=0.6\textwidth]{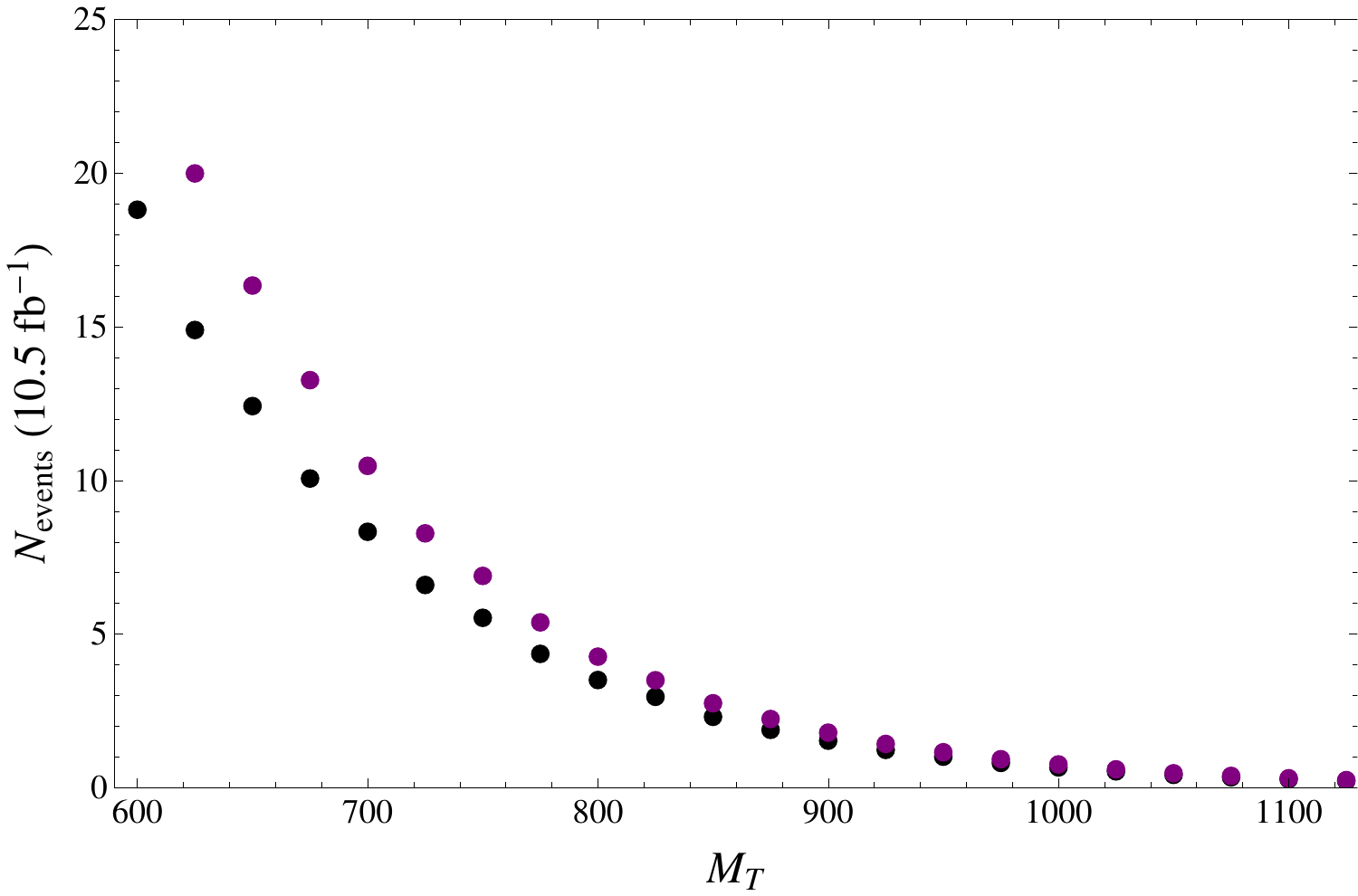}
\caption{Comparison between the signal event yield in the CMS analysis
  (black) and with reduced jet $p_T$ cuts (purple), which could allow to extend the limit by $\unit{50}{\giga\electronvolt}$}
\label{cms_vs_us}
\end{figure}

The analysis we rely on to set bounds on our model, uses very stringent cuts, which reduce the background tremendously but also 
eliminate a significant part of the signal. Among the most restrictive cuts, the requirement for all jets to have a $p_T$ above
$\unit{40}{\giga\electronvolt}$ seems to be rather difficult to pass so we considered an alternative selection where only the two leading
$b$-jets are asked to fulfil this condition, while other jets need only have a $p_T$ bigger than $\unit{20}{\giga\electronvolt}$. As shown in 
Figure~\ref{cms_vs_us}, this increases the event yield of our signal for the lower $M_T$ region but does not provide a significant 
improvement for higher masses. 
Given the fact that the background estimates are data-driven, it is hard to extrapolate it to selections that have not been studied by CMS. 
One could however hope to balance the increase due to this less constraining condition by imposing harder $H_T$ cuts,
which, as shown in Figure~\ref{a} could be as high as $\unit{500}{\giga\electronvolt}$ without reducing the signal in a
noticeable way. For want of a more precise description in the expected background in this case, we predict that such a change in the cuts
would account for an increase of around $\unit{40}{\giga\electronvolt}$ in the hypothesis that
the balance between the change in $p_T$ and $H_T$ cuts is exact. This should not, of course, be seen as an accurate prediction 
but rather as a rough estimate to illustrate this discussion.

\subsubsection{Adding an extra lepton}
Among the main issues one might encounter for such an analysis is the technical difficulty of identifying and tagging many jets in the very 
intense hadronic activity that this kind of events produce. To anticipate this possible limitation, we propose a different analysis (3SSL) 
where the hadronic activity is only controlled with a transverse energy cut, without requiring jet identification. This of course is much 
less constraining than the previous selection relying on the presence of three
$b$-tagged jets so we propose to balance the background increase by increasing the requested isolated 
same-sign lepton count by one, which will drastically reduce any
Standard Model contribution. This is however also harmful for the signal as we now pay the price of another leptonic top decay.

Our selection relies then on the following criteria:
\begin{itemize}
\item 3 leptons of the same sign with $p_T>\unit{20}{\giga\electronvolt}$
\item the sum of the hadronic $p_T$ within a cone $\Delta R<0.3$ of each lepton should not be larger than $10\%$ of the lepton's $p_T$
\item ${\not}E_T>\unit{50}{\giga\electronvolt}$
\item $H_T>\unit{200}{\giga\electronvolt}$
\item $H_{T,had}>\unit{150}{\giga\electronvolt}$
\end{itemize}
In order to investigate the possible reach of such a search, we model the background for
having three same-sign leptons at the LHC using MadGraph and Pythia. This background can be sorted in three categories, depending
on whether the leptons come from a physical process, lepton charge misidentification or fakes from heavy flavour decays. Real three
same-sign lepton processes in the Standard Model are really suppressed since each lepton will come from a separate $W$, $Z$ or $t$ 
decay. 

Table \ref{bkg_sm} summarises the dominating channels and the associated cross-sections at both $\unit{8}{\tera\electronvolt}$ and
$\unit{14}{\tera\electronvolt}$, and shows how small their contribution is.
\begin{table}[!ht]
\centering
\begin{tabular}{|c|c|}
\hline
Channel & $\sigma \times BR\ (\unit{\hspace{-0.2em}}{\pico\barn})$ \\\hline
$ZZW$ & $1.05 \times 10^{-5}$\\\hline 
$ZZW$ & $5.57 \times 10^{-6}$ \\\hline
$ZWt$ & $1.79\times 10^{-6}$\\\hline
$ZWW$ & $6.28 \times 10^{-7}$ \\\hline
$ZZt$ & $4.8 \times 10^{-7}$ \\\hline
$Wtt$ & $1.83 \times 10^{-8}$ \\\hline
$Ztt$ & $1.02 \times 10^{-8}$ \\\hline
$WWW$ & $6.89 \times 10^{-9}$ \\\hline
\end{tabular}
\caption{Standard Model Background to 3SSL. The only kinematic cut
  imposed is $p_{T,l}>\unit{10}{\giga \electronvolt}$ }
\label{bkg_sm}
\end{table}

The contribution of lepton charge misidentification was estimated at $\unit{8}{\tera\electronvolt}$ using an CMS search for
trileptons \cite{Chatrchyan:2014aea}. We take their measured number of events with no hadronic
$\tau$ decay with $H_T>\unit{200}{\giga\electronvolt}$ and ${\not}E_T>\unit{50}{\giga\electronvolt}$, which is $218$
events with $\unit{19.5}{\femto\barn^-1}$ at $\unit{8}{\tera\electronvolt}$ and multiply it by the highest estimation for the lepton charge 
misidentification rate: $10^{-3}$, which results in a cross section contribution of $\unit{1.1\times 10^{-2}}{\femto\barn}$.

Finally, the background from fake leptons stemming from heavy-flavour decay is modelled using the rule-of-thumb presented 
in \cite{Sullivan2010}, that 1 in 200 $b$--quarks typically gives an isolated electron or muon. We show in Table \ref{bkgfake} the
main channels where a $b$--quark decay is at the origin of one of the three leptons. Given that a rather high cross-section of $\approx
\unit{10^{-4}}{\pico\barn}$ is reached in the first three channels under consideration, we studied them in a more precise analysis in 
MG5+Pythia.
\begin{table}[!ht]
\centering
\begin{tabular}{|c|c|}
\hline
Channel & $\sigma \times BR\ (\unit{\hspace{-0.2em}}{\pico\barn})$\\\hline
$Zbb$ & $8.29\times 10^{-4}$\\\hline
$t(b)b$ & $3.04\times 10^{-4}$ \\\hline
$Wt(b)$ & $1.35\times 10^{-4}$ \\\hline
$Zt(b)$ & $8.6\times 10^{-5}$\\\hline
$Wbb$ & $3.12\times 10^{-5}$\\\hline
$Zbt$ & $1.14\times 10^{-6}$\\\hline
$WWb$ & $3.67\times 10^{-8}$\\\hline
\end{tabular}
\caption{Processes with heavy flavour decay contributing to 3SSL. The
  only kinematic cut imposed was $p_{T,l}>\unit{10}{\giga\electronvolt}$. The
  branching $b\to$ isolated lepton is taken as 2\% as advised
  in \cite{Sullivan2010} and for the most computer resource intensive channels,
  the production cross-section is multiplied by appropriate factors of $Z\rightarrow l$
  branching ratios ($\approx 7\%$)
  and $W\rightarrow l$ branching ratios ($\approx 20\%$). The first
  three channels are intense enough to deserve a complete analysis as
  described below.}
\label{bkgfake}
\end{table}

The analysis included the full production of the final state with decayed vector bosons and top quarks at the MG5 level, with further
decay, ISR, FSR and hadronisation in Pythia. Applying the full 3SSL analysis shows that the real background is much lower than the 
estimate, so low in fact that we were unable to generate events passing all the cuts in two channels, for which we give only upper bounds 
for the cross-section. This can be probably explained by the fact that among the two leptons from $B$ decay which are isolated from their 
mother jet, only a small fraction are isolated from the surrounding intense hadronic activity. 
\begin{table}[ht]
\centering
\begin{tabular}{|c|c|}
\hline
Channel & $\sigma \times BR\ (\unit{\hspace{-0.2em}}{\pico\barn})$\\\hline
$Zbb$ & $<3.12\times 10^{-7}$\\\hline
$t(b)b$ & $<1.28\times 10^{-6}$ \\\hline
$Wt(b)$ & $4.98\times 10^{-7}$ \\\hline
\end{tabular}
\caption{Event yield expected with the restricted cuts.}
\label{bkgfakefull}
\end{table}

\begin{figure}[ht]
\centering
\includegraphics[width=0.6\textwidth]{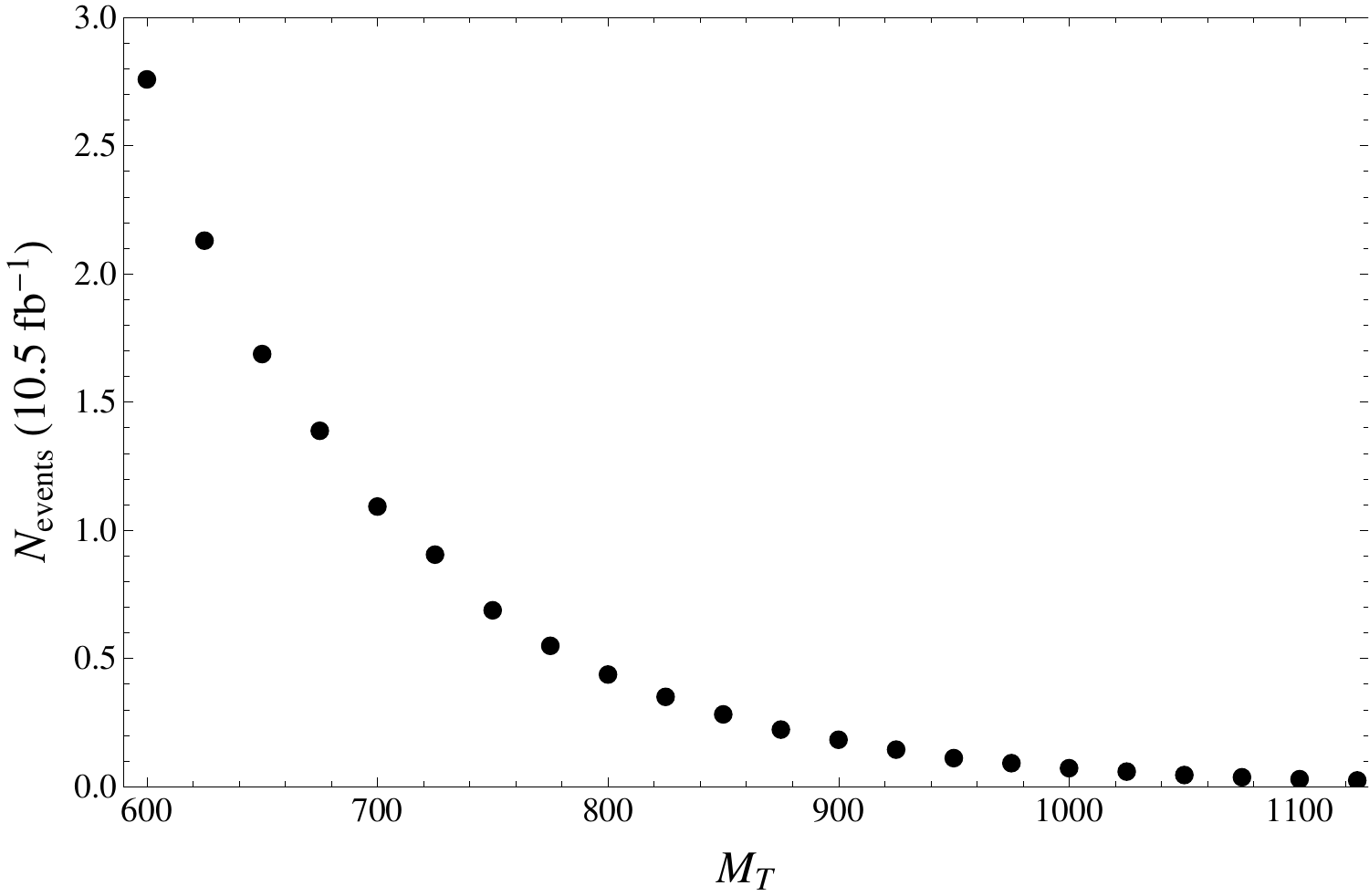}
\caption{Signal event yield expected for a 3SSL analysis at $\unit{8}{\tera\electronvolt}$ with $\unit{10.5}{\femto\barn^{-1}}$ in a scan 
over $M_T$ and $M_Z'=\unit{400}{\giga\electronvolt}$}
\label{mt8}
\end{figure}

\begin{figure}[ht]
\centering
\includegraphics[width=0.45\textwidth]{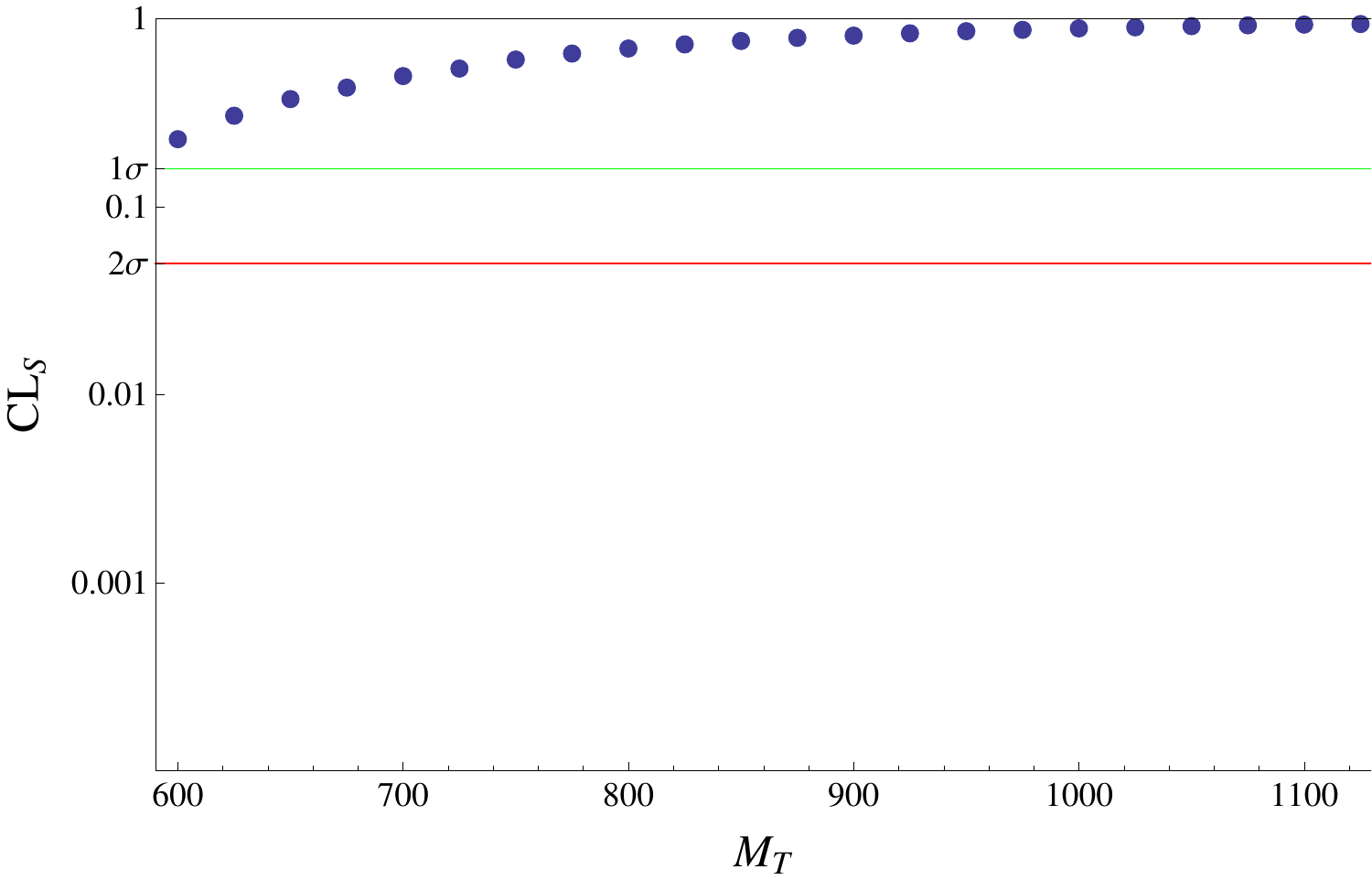}
\includegraphics[width=0.45\textwidth]{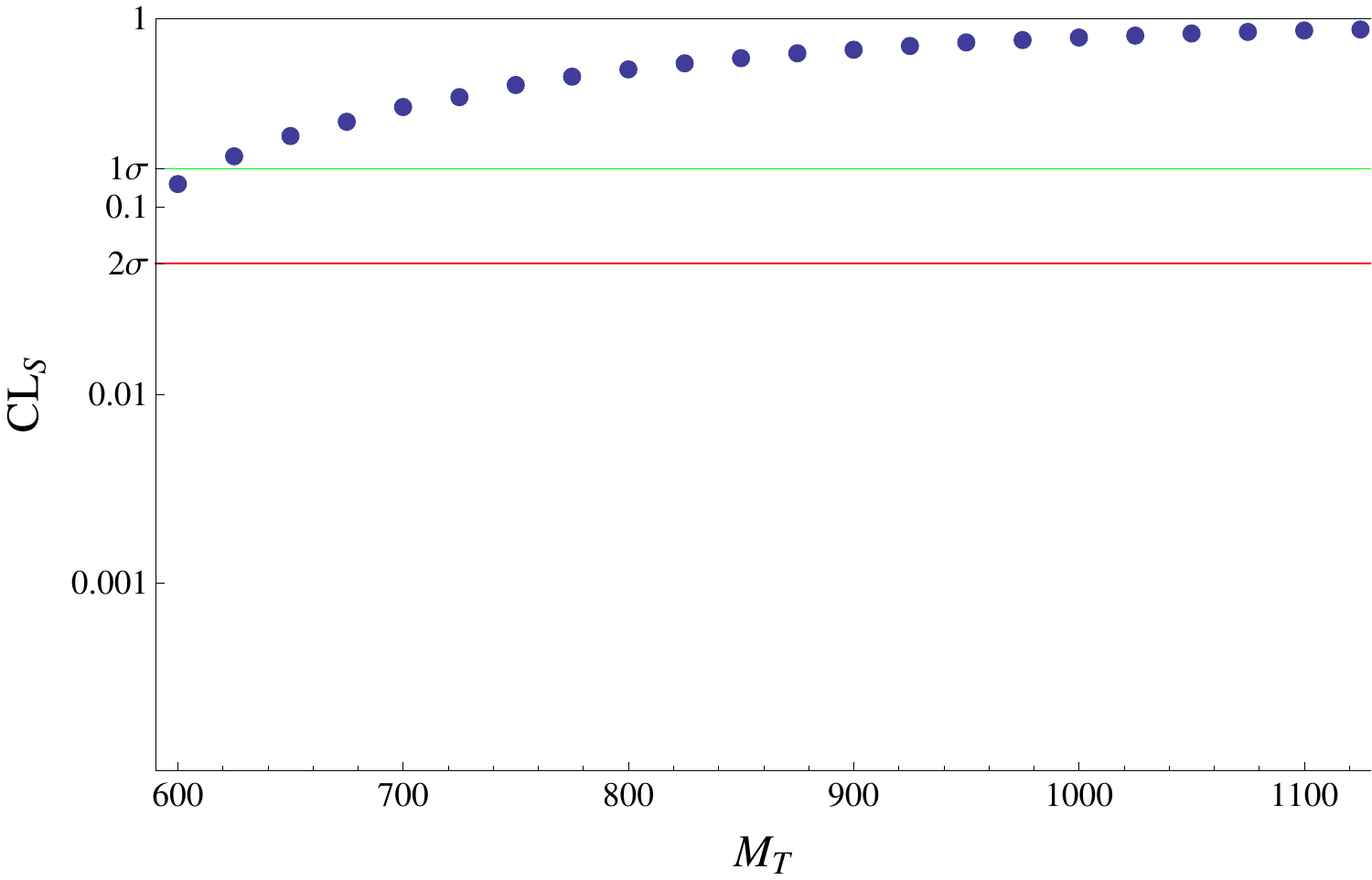}
\caption{Expected confidence level for the 3SSL analysis with the naive, most conservative background estimation (left) and with a 
$10\%$ acceptance factor in simulated channels (right).}
\label{cl8}
\end{figure}

Now that we have some handle on the background we can turn to the signal at $\unit{8}{\tera\electronvolt}$. The number of signal events 
in a scan over $M_T$ is shown in Figure~\ref{mt8}, where it is clear that this channel is less intense than the 2SSL+b channel. If the 
latter should fail due to the reasons we listed above, an analysis using 3SSL might put a weaker limit on our toy model. 
At $\unit{8}{\tera \electronvolt}$, the signal is however so low that with our background estimate that sums up to 
$\approx\unit{1\times 10^{-1}}{\femto\barn}$, the expected limit is below the kinematic limit put on $M_T$, as shown in 
Figure~\ref{cl8} (left panel). Our background estimation is however very conservative so that we display in Figure~\ref{cl8} (right panel) the 
expected confidence level on the exclusion of our model where we add a $10\%$ acceptance to all simulated channels that were not 
processed through the whole 3SSL analysis. We keep the data-based charged misidentification contribution unchanged because the 
CMS analysis already performs cuts very similar to ours. This $10\%$ acceptance factor seems reasonable given the $<10^{-2}$ factor 
between the naive and realistic simulations of the heavy flavour decay contribution. In that case we can set a $1\sigma$ 
limit at $\unit{625}{\giga\electronvolt}$, which is a little above the kinematic limit. 
We shall however see in the next section that there is hope 
to increase this limit by a large value with the LHC at $\unit{14}{\tera\electronvolt}$.

\subsubsection{Four top processes}
While this model was constructed to study six top final states, it can yield four tops as well in all color embedding. Indeed, a highly
off--shell top in a pair-production might radiate a $Z'$ whose decay gives two additional quarks. This channel, however, is
negligible, as we checked by comparing its contribution to 2SSL+b to that of the 6 top process, which is indeed largely dominant by two
orders of magnitudes.
\begin{figure}[ht]
\centering
\includegraphics[scale=1]{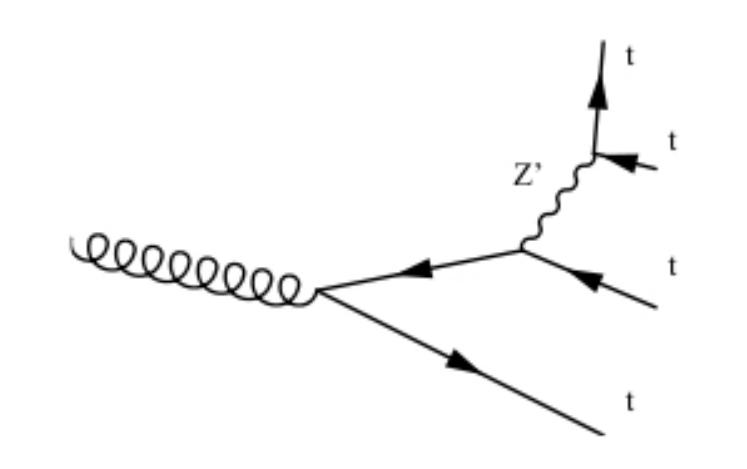}
\label{tttbar}
\caption{Four top production by $Z'$-Strahlung}
\end{figure}
A significant production rate can however be achieved when the $Z'$ is a coloured (coloron-like) vector. The QCD coloured $Z'$ 
pair-production is by far more important compared to $T$ pair-production as we only consider the
region of phase space where $M_T>M_{Z'}+m_t$. The limit set by this process will be much more stringent than those set by the six-top
channel, which means that large-multiplicity final states are not of great interest when the $Z'$ is colored, and we will set it aside in 
the rest of this work.

\subsection{Perspectives at 14 TeV}
\begin{figure}[ht]
\centering
\includegraphics[width=0.6\textwidth]{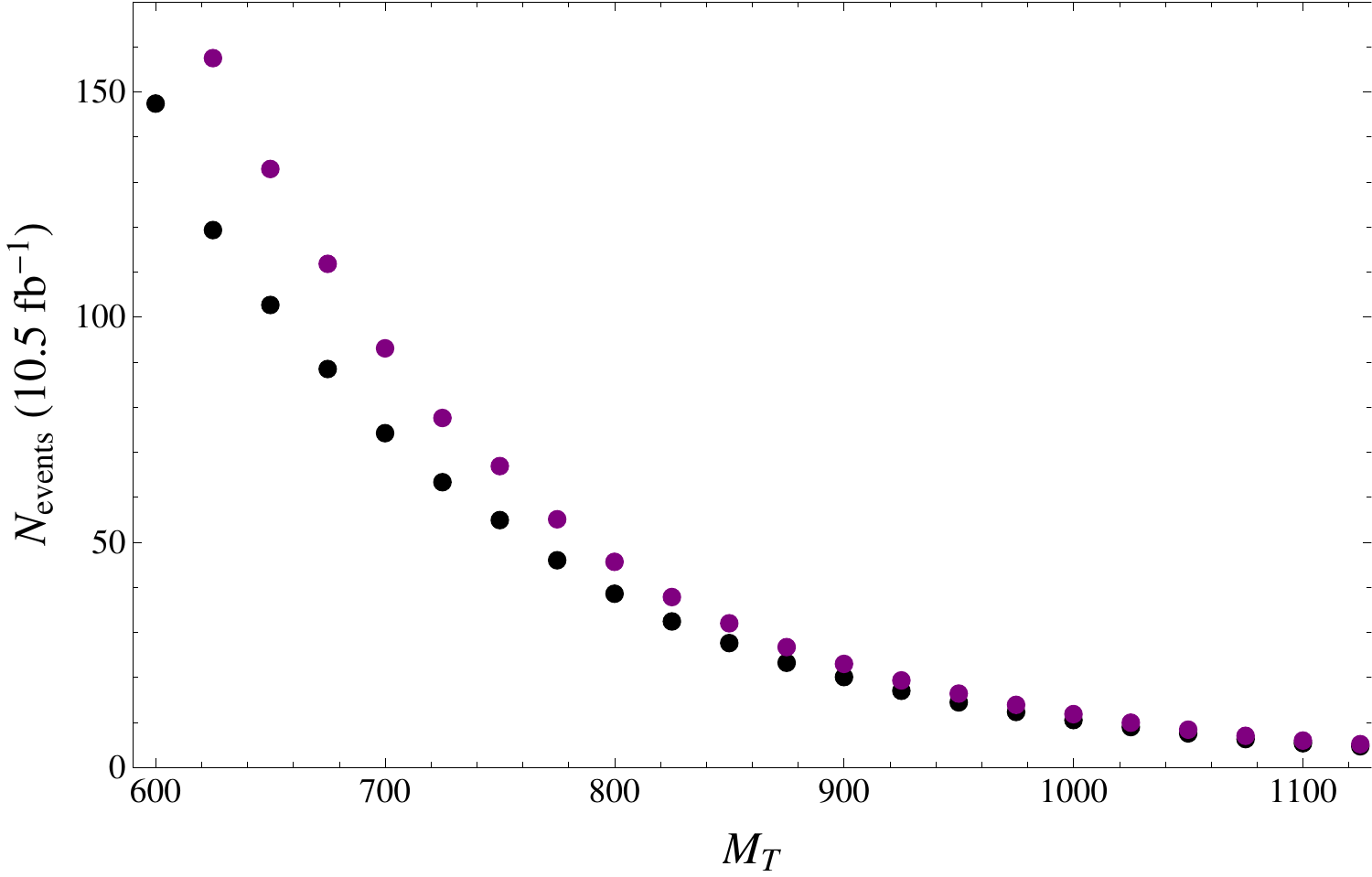} 
\caption{Expected number of signal events in a 2SSL+b with normal (black) and reduced (purple) jet $p_T$ cuts
  analysis at $\unit{14}{\tera\electronvolt}$ in a scan over $M_T$ with
$M_{Z'}=\unit{400}{\giga\electronvolt}$ for
$\unit{10}{\femto\barn^{-1}}$}
\label{predictmta}
\end{figure}

The analysis used to constrain our model has a background which is very hard to properly reproduce in simulations, which is
why experiments do not rely on Monte Carlo samples but use data-driven techniques to estimate it. This, however makes it 
impossible to scale their results up to higher energies to precisely establish the expected
limits for the next run of the LHC. We can however rely on the smallness of the background to make gross estimate of the observation
window for a given amount of data, knowing that several events would probably be
sufficient for having an observable signal. Figure~\ref{predictmta} shows the
event yield expected for a 2SSL+b analysis with $\unit{10.5}{\femto\barn}$, which shows that the reach could be enhanced rather 
strongly if the background does not increase too much. 
One can also project the expected yield for the three same-sign lepton analysis at $\unit{14}{\tera\electronvolt}$. As for the previous case,
it is impossible to make a sensible extrapolation of the background because part of it is data-based. We can hence only present the event
yields as a function of $M_T$. As was shown already at $\unit{8}{\tera\electronvolt}$, the cross-section is significantly
decreased compared to the 2SSL+b case but gains a very large factor compared to the $\unit{8}{\tera\electronvolt}$ case, which 
means that one could hope for this channel to start being competitive, since its background will most likely not scale as fast with energy.
\begin{figure}[ht]
\centering
\includegraphics[width=0.6\textwidth]{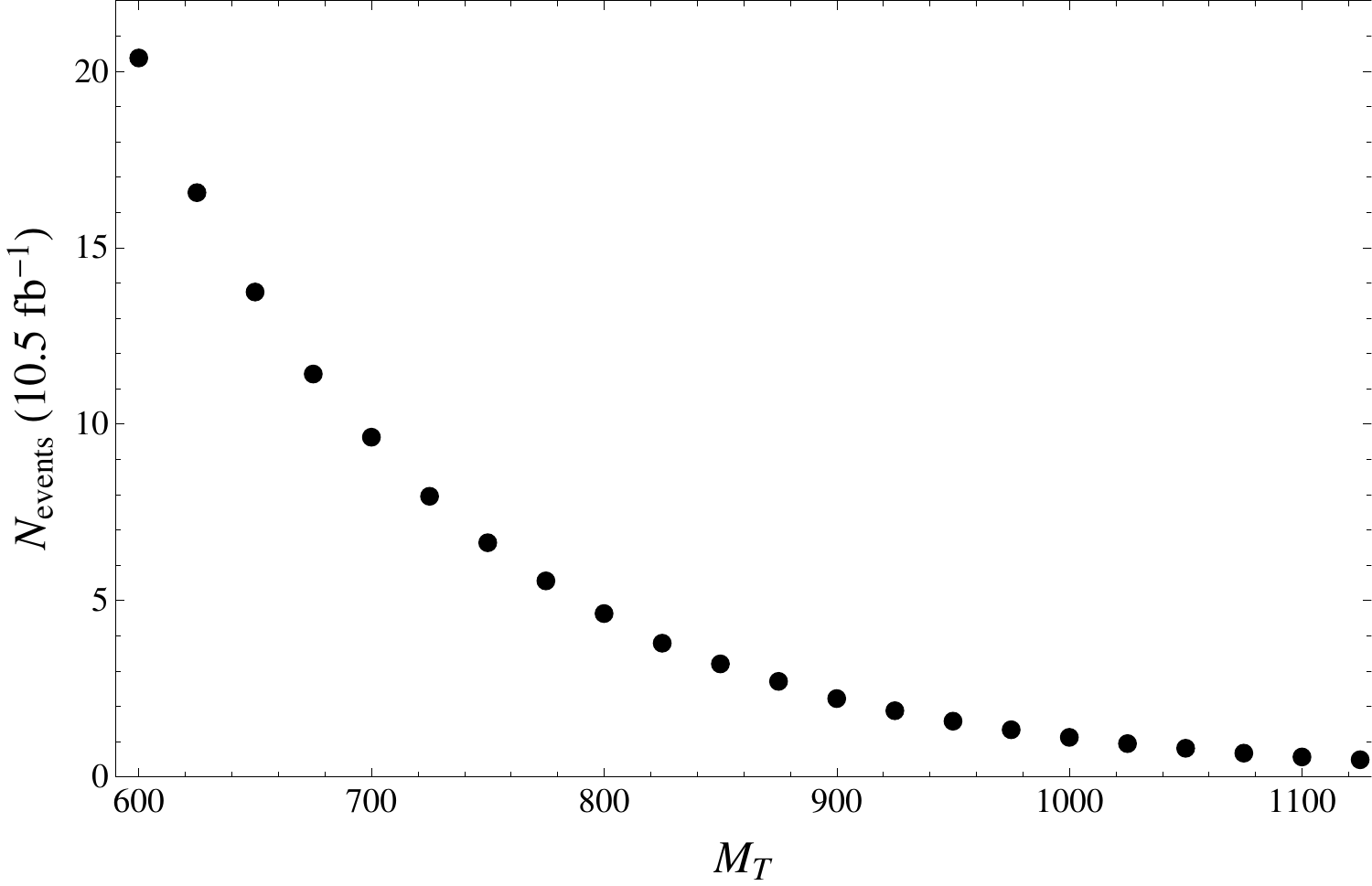}
\caption{Number of signal events expected for a 3SSL analysis at
  $\unit{14}{\tera\electronvolt}$, in a scan over $M_T$ with
  $M_{Z'}=\unit{400}{\giga\electronvolt}$,  with
$\unit{10.5}{\femto\barn^{-1}}$}
\label{3SSL14}
\end{figure}

\section{Eight tops}

\subsection{Model and production process}
It is easy to build upon our first toy model to have a decay chain producing eight top quarks in the final state by adding an extra bosonic 
coloured particle $\rho$ decaying to $t\bar{T}$. This resonance can be pair produced in $pp$ collisions by QCD and it can give rise to 
the expected signature. In this model, the six-top final state can obviously also arise and all the above discussion applies, but we want to 
investigate the phenomenology of the eight top final state.

The colour assignments for $T$ and $Z'$ are constrained exactly as before, which imposes $\rho$ to be a colour-octet if we stick to the 
minimal and phenomenologically more reasonable choice for $T$ and $Z'$. For the sake of the discussion, we will also set it as a real 
vector field.

\subsection{Phenomenology of the eight-top final state}
\begin{figure}[ht]
\centering
\subfigure[\hspace{-1em}]{\includegraphics[scale=0.45]{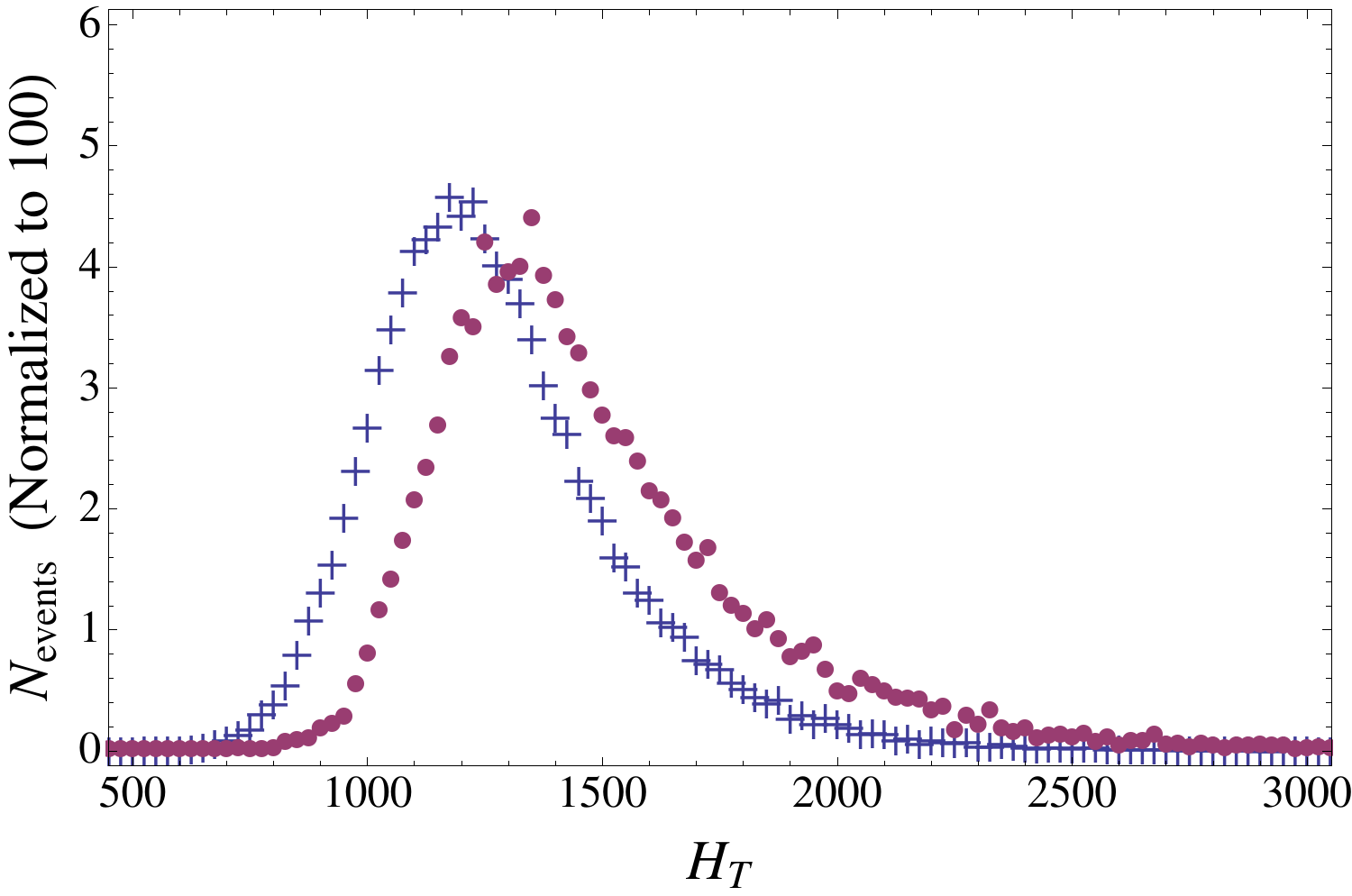}\label{86a}}\hspace{0.9em}
\subfigure[\hspace{-1em}]{\includegraphics[scale=0.45]{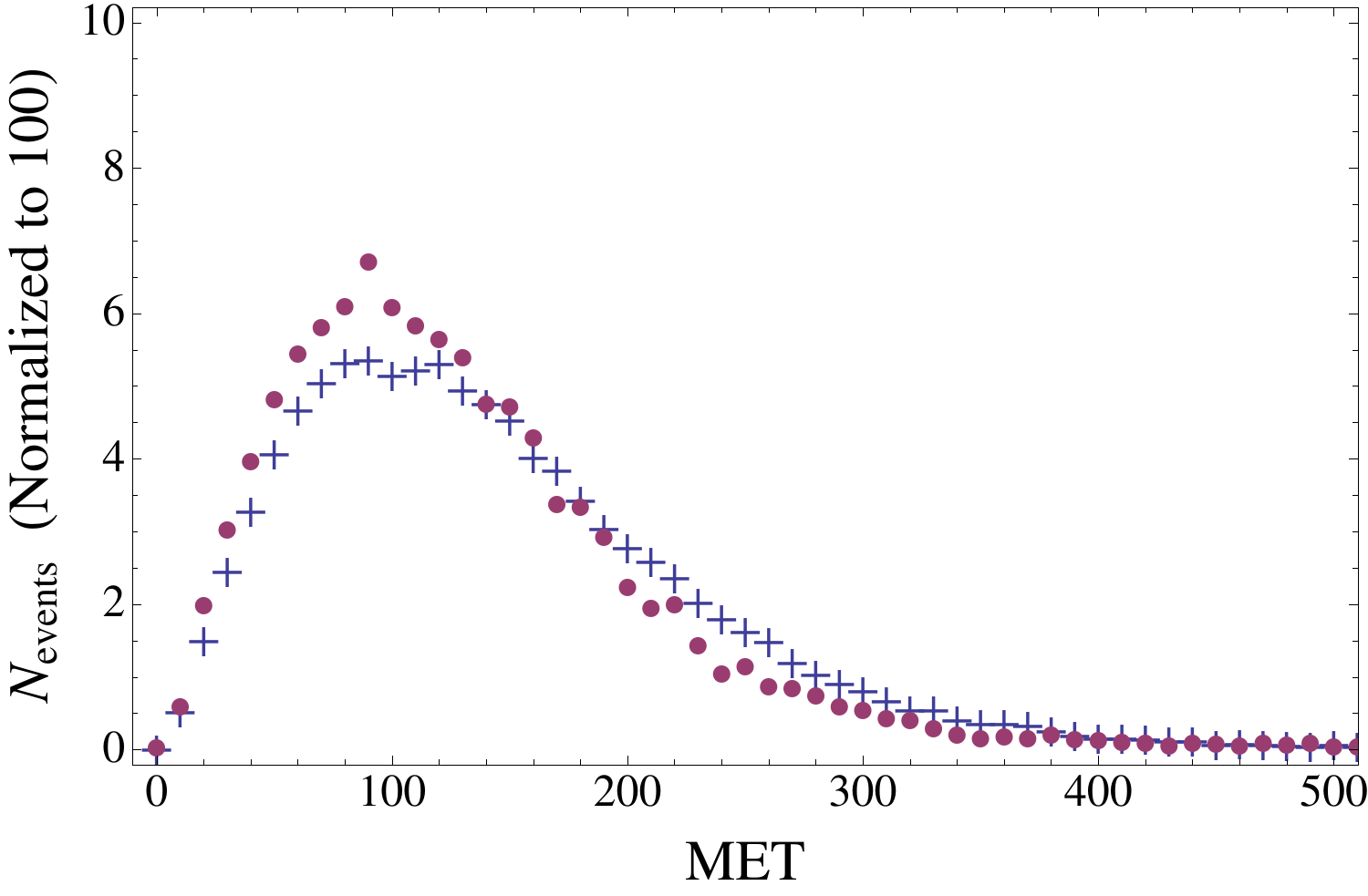}\label{86b}}\\
\subfigure[Hardest lepton]{\includegraphics[scale=0.45]{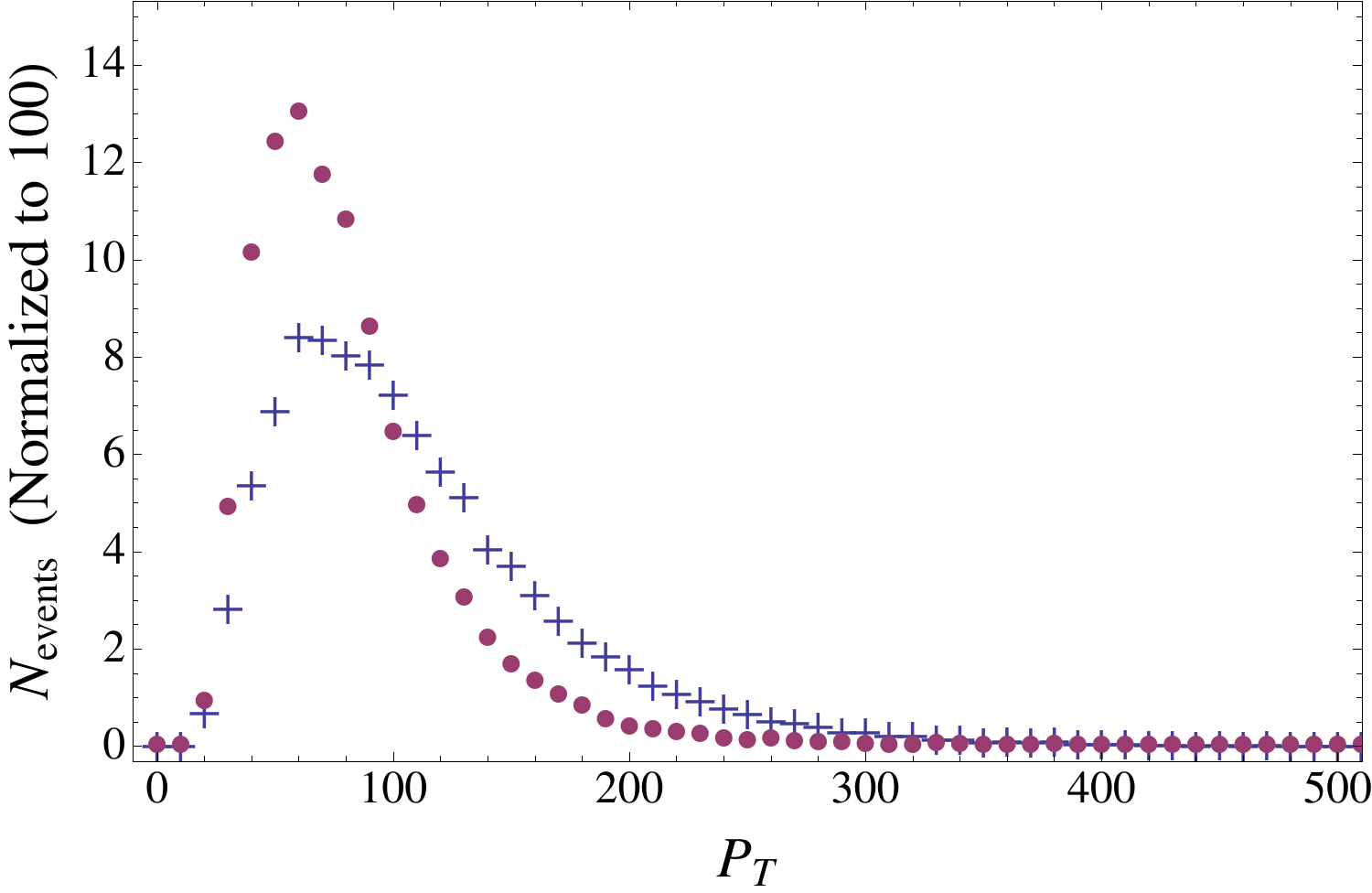}\label{86c}}\hspace{0.9em}
\subfigure[\hspace{-1em}]{\includegraphics[scale=0.45]{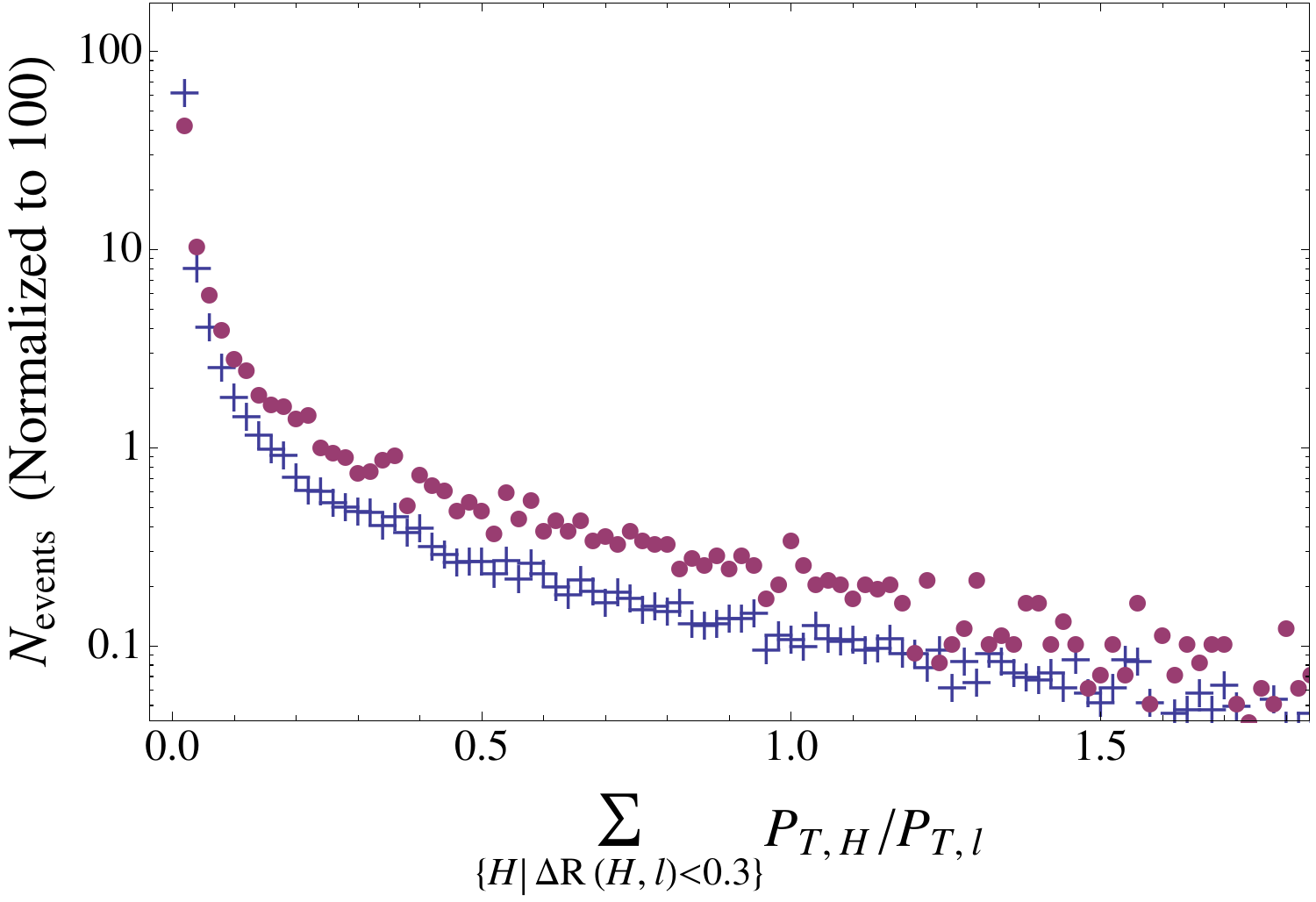}\label{86d}}\\
\caption{Kinematic distributions of the decay products of a pair of $\unit{800}{\giga\electronvolt}$ $\rho$ (purple dots) and $T$ 
(blue crosses) at $\unit{8}{\tera\electronvolt}$.}
\label{8vs6}
\end{figure}

The most promising ways to detect a multi-top final is through a leptonic channel. As before, exploiting the possibility of having same-sign 
leptons is probably the best way to reduce Standard Model backgrounds. It is therefore useful for understanding how the dynamics differs 
from the six-top case, as presented in Figure~\ref{8vs6}.
One can see that as expected, the $H_T$ spectrum is harder for eight-top events. The ${\not}E_T$ and leptonic $p_T$ are slightly lower 
than in the six-top case, which is understandable since individual tops have lower momentum. The extra hadronic activity is also visible 
in the distribution of the leptonic isolation variable (sum of the hadronic $p_T$ in the vicinity of a lepton divided by the lepton's $p_T$), 
which means that the leptonic cuts have a lower acceptance in the eight-top final state. These factors combine to eliminate the 
possibility to put any bound on $M_{\rho}$ with the 2SSL+b analysis using current data, as for a $\unit{800}{\giga\electronvolt}$, the 
results of the CMS analysis is compatible with the background only hypothesis within less than $1\sigma$. The 3SSL analysis is no 
better off: even though combinatorics help reducing the signal loss due to the third branching ratio to leptons, compared to the six-top 
case, no signal event is expected at the LHC with the current amount of data at $\unit{800}{\giga\electronvolt}$.

\section{Conclusions}

This article tackles in a preliminary way the question of the maximal top multiplicity that could be detected in order to motivate searches 
for those final states that could be accessible. In a very broad analysis based on toy models, we could establish that current LHC data 
could allow to place a limit on new physics giving rise to six-top final states, but that eight-top final states are beyond our reach. We could 
also show that there is hope to increase the window in top multiplicity once the LHC runs at nominal energy, as much stronger bounds 
should be placed on six-top processes, and eight top processes should begin to become visible. This work is however more of a proof-of-
principle than a precise analysis, which should be conducted in a more realistic setup by the experimental collaborations, in order to be 
able to place precise bounds. 

\subsection*{Acknowledgements}
We thank G. Cacciapaglia for participation to the early stage of this project and for numerous discussions. We also thank R. Chierici and F. Maltoni for helpful discussions and comments. We thank M. Zaro for his help in making Figure \ref{plotsm} and with MG5, and B. Fuks for his help with MA5.
AD is partially supported by Institut Universitaire de France. We also acknowledge partial support from 
the Labex-LIO (Lyon Institute of Origins) under grant ANR-10-LABX-66 and FRAMA (FR3127, F\'ed\'eration de Recherche ``Andr\'e 
Marie Amp\`ere").
  
\end{document}